\documentclass[referee]{aa} 
\usepackage{graphicx}
\begin{document}
\title{Abundances for metal-poor stars with accurate parallaxes
\thanks{Based in part on data collected at the European Southern
Observatory, Chile, at the MacDonald Observatory, Texas, USA, and at the
Telescopio Nazionale Galileo, Canary Island, INAF, Italy-Spain}}
\subtitle{I. Basic data}

\author{R.G. Gratton\inst{1}, 
E. Carretta\inst{1}, 
R. Claudi\inst{1}, 
S. Lucatello\inst{1,2},
\and
M. Barbieri\inst{3}}

\offprints{R.G. Gratton}

\institute{INAF - Osservatorio Astronomico di Padova, Vicolo dell'Osservatorio
5, 35122 Padova, Italy\\
\and
Dipartimento di Astronomia, Universit\`a di Padova, Italy
\and
CISAS, Universit\`a di Padova, Italy}

\date{Received ; accepted}

\abstract{
We present element-to-element abundance ratios measured from high dispersion
spectra for 150 field subdwarfs and early subgiants with accurate Hipparcos
parallaxes (errors $<20$\%). For 50 stars new spectra were obtained with the
UVES on Kueyen (VLT UT2), the McDonald 2.7m telescope, and SARG at TNG.
Additionally, literature equivalent widths were taken from the works by Nissen
\& Schuster, Fulbright, and Prochaska et al. to complement our data. The whole
sample includes both thick disk and halo stars (and a few thin disk stars);
most stars have metallicities in the range -2$<$[Fe/H]$<$-0.6. We found our
data, that of Nissen \& Schuster, and that of Prochaska to be of comparable
quality; results from Fulbright scatter a bit more, but they are
still of very good quality and are extremely useful due to the large size of
his sample. The results of the present analysis will be used in forthcoming
papers to discuss the chemical properties of the dissipational collapse and
accretion components of our Galaxy
   }

\authorrunning{Gratton R.G. et al.}
\titlerunning{Abundances in metal poor-stars}

   \maketitle
%

\section{Introduction}

An accurate comparison between the abundances measured from spectra of stars
in globular clusters and field stars is important when considering various
issues:
\begin{itemize}
\item Determination of distances to globular cluster stars using the main
sequence fitting method (Reid 1997; Gratton et al. 1997; Pont et al. 1998; Reid
\& Gizis 1998; Carretta et al. 2000) requires accurate and homogeneous
abundances for field and cluster stars: a systematic offset of 0.1 dex between
the two sets of abundances yields an error of 0.08 mag in the distance scale,
and of $\sim 1$~Gyr in the ages
\item The epoch and formation mechanism of globular clusters and field stars
may be constrained by the abundance ratios for key-elements. In order to avoid
ambiguities, observation of several elements is required. For instance, a low
(roughly solar) ratio between the abundances of $\alpha-$elements and Fe might
indicate a significant contribution to nucleosynthesis by thermonuclear SNe:
however a possible alternative interpretation is that only small mass
core-collapse SNe (which likely do not efficiently produce $\alpha-$elements)
are involved. A solution of this ambiguity may be obtained by observations of
elements (like e.g. Mn) that are produced in the same layers of the
progenitors of core-collapse SNe where Fe is produced
\item A detailed discussion of the formation mechanisms of metal-poor stars
requires not only a knowledge of the runs of average abundance ratios, but also
possibly determination of intrinsic spreads around them. This is a very
valuable information than can be used to constrain the size of clouds
undergoing independent chemical evolution. In fact, if such clouds are not
extremely large, we may expect to observe an intrinsic star-to-star scatter
related to the random sampling of the initial mass function for the
progenitors of the core-collapse supernovae (see e.g. discussion in Carretta
et al. 2002). This requires that observations should be both accurate and
extended over large samples. Furthermore, it would be very useful to discuss
such runs and dispersions around them for different galactic populations
(mainly halo and thick disk); this requires knowledge of the star kinematics,
which on turn needs a relatively good knowledge of stellar distances (although
in this case the distance errors that can be accepted - $\Delta \pi/\pi<0.2$\
- are somewhat larger than for the determination of distance scales).
\end{itemize}

Thus, observation of an extended sample of metal-poor field stars with accurate
parallaxes is of the highest importance. For the first purpose it is
important that temperatures for the field stars are derived using the same
procedure used for globular clusters; it is also important that these
temperature determinations are based on reddening free parameters, because
there might be systematic offsets between the reddening scale usually adopted
for (far) globular clusters and (nearby) field stars, due to the current
uncertainties in the dust layer scale-height. An uncertainty of 0.01 mag in
the relative reddening scales for globular clusters and local subdwarfs may
in fact yields an error of 0.08 mag in the distance scales, and of 1 Gyr in 
the ages.

For this reason, we decided to include in the ESO Large Program 165.L-0263
also observations of a number of subdwarfs with accurate Hipparcos parallaxes
and metallicities [Fe/H]$<-0.6$\footnote{In this paper we will use two distinct
notations for the abundances of elements in stellar atmospheres: $\log n({\rm
A})$\ is the abundance of the element A (by number) in a scale where the
abundance of H is 12; and [A/B] is the difference between the logarithms of
the abundance ratio of two elements in a star and in the Sun: [A/B]$\log
n({\rm A/B})-\log n({\rm A/B})_\odot$}. These stars were observed using the
same set up devised for Turn-off (TO) stars and subgiants in globular clusters,
so that homogeneous temperature scales could be derived. For the two other
purposes mentioned above, this core sample of field metal-poor stars may be
greatly enlarged by considering other sets of high quality equivalent widths
available. We first considered spectra of 12 stars taken by us for other
programs (Gratton et al. 2000a) at the McDonald Observatory. We then added the
high quality data for about 30 thick disk and halo stars from Nissen \&
Schuster (1997), those for thick disk stars by Prochaska et al. (2000), and
finally those from the large survey of metal-poor stars by Fulbright (2000).
To maintain homogeneity throughout our study, we only considered main sequence
and early subgiant stars. In fact, within our analysis, we are mainly
interested in relative abundances: since the effects we wish to show are not
very large, adoption of an homogeneous set of atmospheric parameters (mainly
effective temperatures) for field and cluster stars is basic. Such homogeneous
analysis is much easier for TO-stars and early subgiants, because for these
stars we may obtain quite accurate temperatures and gravities.

As to field stars, our final aim is to obtain homogeneous abundances for all
those metal-poor stars with accurate parallaxes from Hipparcos
($\Delta\pi/\pi<0.12$) included in the {\it a priori} sample devised by
Carretta et al. (2000) \footnote{It should be noted that two of the stars
considered by Carretta et al. (2000) - HD6755 and HD17072 - are evolved
giants, the second being a red horizontal branch stars (Gratton 1998; Carney
et al. 1998). Also, BD-0~4234 is a cool BY~Dra variable, for which abundance
analysis is unreliable. These stars will be not further considered in our
papers. Hence, the total sample of useful stars from Carretta et al. (2000)
is composed of 51 stars} and of a large number of those metal-poor stars
([Fe/H]$<-0.5$) with errors in parallaxes $\Delta\pi/\pi<0.2$. In this paper
we present the analysis of the abundances of $\alpha-$ and Fe-group elements
for a group of 150 stars (including 48 of the stars of Carretta et al. 2000).
Some of the data here considered have been already presented in a previous
paper (Gratton et al. 2001; hereinafter Paper I), where we concentrated
on the star-to-star abundances within globular clusters.

In this first paper of the series on field stars, we will present the basic
data on which our discussion is based, including the stellar photometric and
kinematic data, and the chemical abundances. The second paper will be devoted
to the discussion of the abundance ratios between Fe and the $\alpha-$elements.
Finally, the remaining analyzed elements will be considered in a third paper.

\begin{table}
\caption{Equivalent widths measured on UVES, SARG and McDonald spectra (only
available in electronic form)}
\label{t:ew}
\end{table}

\section{Sample selection and Observations}

Our sample consists of four sets of data:
\begin{itemize}
\item Spectra for 40 field stars were obtained using the UVES spectrograph at
Kueyen (=VLT UT2) telescope on Paranal in various runs between 2000 and 2001.
These spectra have a resolution of $R\sim 50,000$, and a $S/N\sim 200$. They
were obtained using the dichroic beamsplitter \#1, with a slightly different
setup in the first run and in the other runs. However, the covered spectral
range is very broad: blue spectra cover the range 3,600-4,800~\AA, and the red
ones 5,500-9,000~\AA, with essentially no gaps between the orders. The spectra
were reduced using the UVES pipeline. Additionally, spectra for three stars
were obtained with the SARG spectrograph at the Italian TNG telescope at La
Palma. They have a very high resolution of 150,000, and a S/N of $\sim 100$.
The spectral coverage is from about 4,700 to 7,900~\AA. These spectra were
reduced using IRAF routines
\footnote{IRAF is distributed by the National Optical Observatory, which is
operated by the Association of Universities for Research in Astronomy, Inc.,
under contract with the National Science Fundation}
. Equivalent widths ($EW$s) on all these spectra were
measured using an our own automatic procedure, that uses a gaussian fitting
routine (see Bragaglia et al. 2001 for details). 
They are listed in Table~\ref{t:ew}, only available in electronic
form. Accuracy of these $EW$s is very good. From measures on different
spectra that were available for four stars, we obtain standard deviations of
2.1~m\AA\ on individual $EW$s, with no outliers discarded; note however that
the error distribution is not gaussian: for most lines, measures on different
spectra agree better than 1.5~m\AA.

\item Equivalent widths were further measured on the spectra of 12 main 
sequence and early subgiants gathered in the last years with the high 
dispersion spectrograph at the 2.7m telescope of the McDonald Observatory;
these spectra were used in Gratton et al. (2000a). They have a resolution of
80,000, a S/N$\sim$200, and a very broad spectral coverage (from about 
3,700 to about 9,000~\AA). $EW$s were measured with an interactive Gaussian 
fitting routine. This procedure is slightly less accurate than that used for
the UVES spectra (mainly due to the way the reference continuum is set); this
results in slightly larger errors in the $EW$s (about $\pm 3$~m\AA), as given
by a line-to-line comparison with UVES data for two stars in common (HD134439
and HD194598).

\item In addition to this original set of data for 50 stars, we also considered
recent high quality literature analysis of similar stars. Nissen \& Schuster
(1997) presented a careful analysis of abundances in 13 halo and 16 disk
subdwarfs and early subgiants, with metal abundances in the range
$-1.3\leq$[Fe/H]$\leq-0.5$. The observational material consisted in spectra
taken at a resolution $R=60,000$\ and $S/N\sim 150$\ with the EMMI
spectrograph at the ESO NTT telescope. We considered here 23 stars that have
parallaxes measured with accuracy $\Delta\pi/\pi<0.2$. 

\begin{figure}
\centering
\includegraphics[width=8.8cm]{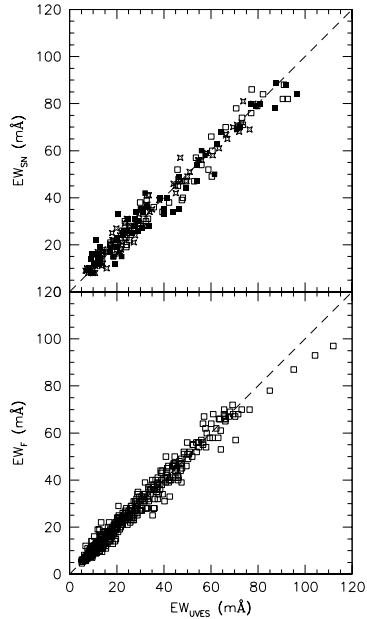}
\caption{ Comparison between $EW$s measured from the UVES spectra and those from
other sources for the same stars. Upper panel: $EW$s from Nissen \&
Schuster (1997); lower panel: $EW$s from Fulbright (2000) }
\label{f:compew}
\end{figure}  

A graphic comparison between the $EW$s measured on our spectra and by Nissen \&
Schuster (1997) for the three stars in common is shown in the upper panel of
Fig.~\ref{f:compew}. There is no significant systematic offset for any of
the three stars. On average, $EW$s measured on our spectra are smaller by
$0.4\pm 0.3$~m\AA\ (192 lines; r.m.s. scatter of 4.4~m\AA); if 12 lines for
which differences are larger than 2~$\sigma$\ are not considered, the mean
difference is $0.1\pm 0.3$~m\AA\ (180 lines; r.m.s. scatter of 3.7~m\AA). The
$EW$s measured by Nissen \& Schuster seem then of an accuracy comparable to
those of our data, consistently with the similar quality of the observational
material.

\item Prochaska et al. (2000) measured abundances in a sample of 10 thick disk
stars using spectra acquired with the HIRES spectrograph at the Keck 10 m
telescope. The spectra have a nearly complete spectral coverage from
4400 to 9000~\AA, with the exception of inter-order gaps. The spectra have
a resolution of $R\sim 50,000$\ and a $S/N>100$. Eight of these stars have
parallax error $\Delta\pi/\pi<0.2$, and are considered in this paper

\item Fulbright (2000) recently published a very extensive analysis of the
abundances of 168 field halo and disk stars. Data were obtained over the
period 1994-1999 with a variety of instruments at different telescopes.
The original spectra have a resolution of $R\sim 50,000$\ and a $S/N>100$;
they have then a quality comparable to that of our UVES spectra. In this
paper, we only considered those stars in his original sample that have
parallax error $\Delta\pi/\pi<0.2$: the stars eliminated in this way are
mainly cool giants. Five other stars having parallax error better than this 
limit were eliminated: HD6755 and HD122563, that are evolved giants;
HD40701 and HD45205 that are binaries, and for which no reliable temperature 
could be determined (see below); and BD+45~983, for which we get a
scatter of abundances from individual lines considerably larger than for
the remaining stars (only a single spectrum with $S/N=90$\ was used by
Fulbright for this star). In total, we considered 108 stars from the
Fulbright sample.

The lower panel of Fig.~\ref{f:compew} gives a comparison between our $EW$s
measured on UVES spectra and those of Fulbright (2000) for the 15 stars in
common. The mean difference is $0.1\pm 0.1$~m\AA\ (430 lines; r.m.s. scatter
of 3.1~m\AA). The agreement between the two sets of $EW$s is excellent, and
suggests that Fulbright $EW$s are approximately as accurate as ours, with
typical errors of $\pm 2$~m\AA\ and no systematic offset. Note that for lines
with $EW>80$~m\AA, there is a trend for our $EW$s to be larger (these lines
are only a tiny fraction of those measured). We think that 
our $EW$s for such lines
might be somewhat underestimated, because they were measured using a gaussian
fitting routine which neglects the damping wings (that are not entirely
negligible for these lines). Since Fulbright $EW$s are even smaller than ours
for these lines, we suspect that he also somewhat underestimated the $EW$s
of the stronger lines.

While the quality of the $EW$s of Fulbright spectra is as good as that of our
$EW$s, it should be noted that he generally considered a somewhat less extended
list of lines for each star; for this reason, his abundances have slightly
larger internal errors (being however still of a very good quality, fully
adequate for the present discussion).

\end{itemize}

\begin{table*}
\caption{Basic parameters. Bin=0: bona fide single star; =1: known or suspected binary}
\scriptsize
\begin{tabular}{rrrrrrrrrrrrrl}
\hline
HD/BD &    RA &     dec &$\mu_{\rm RA}$&$\mu_{\rm dec}$&$V_{\rm rad}$&$\pi$&$\Delta\pi$&$V$&$M_V$&Mass&$B-V$&$b-y$&Bin,\\
         & HHMMSS &  DDMMSS &  mas  &  mas  &  km/s  &  mas   & mas  & mag & mag & $M\odot$&  mag  &  mag  &$E(B-V)$\\       
\hline
  224930 &    210 &  270456 &   779 &  -919 &  -36.2 &  80.63 & 3.03 &  5.75 & 5.28 & 0.78 & 0.670 & 0.432 & 1 \\
    3567 &   3831 &  -81833 &    19 &  -547 &  -49.7 &   9.57 & 1.38 &  9.26 & 4.16 & 0.81 & 0.456 & 0.328 & 0 \\
    3628 &   3913 &   30802 &   781 &   297 &  -28.3 &  21.79 & 0.88 &  7.30 & 3.99 & 0.91 & 0.638 & 0.402 & 1 \\
-35~0360 &  10406 & -344029 &   647 &  -267 &   40.1 &  16.28 & 1.76 & 10.25 & 6.31 & 0.68 & 0.765 & 0.469 & 0 \\
    6582 &  10816 &  545513 &  3421 & -1600 &  -97.2 & 132.42 & 0.60 &  5.12 & 5.73 & 0.75 & 0.694 & 0.437 & 1 \\
    9430 &  13258 &  234144 &  -205 &  -162 &  -53.4 &  15.33 & 1.24 &  9.03 & 4.96 & 0.83 & 0.620 &       & 1 \\
-61~0282 &  13606 & -610503 &   -79 &  -676 &  226.8 &  11.63 & 1.19 & 10.10 & 5.43 & 0.74 & 0.530 & 0.365 & 0 \\
   10607 &  14115 & -674037 &   321 &  -452 &   -0.2 &  14.00 & 0.74 &  8.32 & 4.05 & 0.84 & 0.557 & 0.372 & 1 \\
+29~0366 &  21025 &  294824 &   290 &  -266 &   24.2 &  17.66 & 1.29 &  8.77 & 5.00 & 0.79 & 0.577 & 0.390 & 0 \\
-01~0306 &  21440 &  -11205 &   995 &   -80 &   19.2 &  16.17 & 1.34 &  9.09 & 5.13 & 0.78 & 0.577 & 0.383 & 0 \\
   15096 &  22602 &   54647 &   352 &    83 &  -11.0 &  35.86 & 1.15 &  7.95 & 5.72 & 0.78 & 0.800 & 0.485 & 1 \\
   16397 &  23828 &  304900 &  -488 &  -387 & -100.3 &  27.89 & 1.12 &  7.34 & 4.57 & 0.85 & 0.591 & 0.387 & 0 \\
   17288 &  24410 & -600322 &   357 &    -2 &    8.2 &  10.38 & 1.09 &  9.88 & 4.96 & 0.80 & 0.577 & 0.378 & 0 \\
   17820 &  25158 &  112212 &    37 &  -445 &    6.3 &  15.38 & 1.39 &  8.38 & 4.31 & 0.85 & 0.549 & 0.397 & 0 \\
   18907 &  30138 & -280530 &   283 &  -441 &  -38.4 &  32.94 & 0.72 &  5.85 & 3.44 & 0.88 & 0.790 & 0.495 & 0 \\
   19445 &  30826 &  261952 &  -210 &  -830 & -139.3 &  25.85 & 1.14 &  8.05 & 5.11 & 0.71 & 0.459 & 0.351 & 0 \\
   20512 &  31827 &  151038 &    -1 &  -301 &   10.9 &  17.54 & 1.28 &  7.41 & 3.63 & 0.92 & 0.790 & 0.485 & 1 \\
-47~1087 &  33444 & -471612 &   232 &    68 &   16.2 &   9.28 & 1.34 & 10.26 & 5.10 & 0.79 & 0.570 & 0.387 & 0 \\
   22879 &  34022 &  -31301 &   690 &  -214 &  114.2 &  41.07 & 0.86 &  6.74 & 4.81 & 0.81 & 0.541 & 0.365 & 0 \\
   23439 &  34702 &  412538 &   599 & -1240 &   49.6 &  40.83 & 2.24 &  8.18 & 6.23 & 0.70 & 0.765 & 0.487 & 1 \\
   24616 &  35359 & -230813 &   336 &  -298 &   98.3 &  15.87 & 0.81 &  6.71 & 2.71 & 0.87 & 0.817 & 0.512 & 0 \\
   25704 &  40145 & -571225 &   347 &   414 &   52.0 &  19.02 & 0.87 &  8.10 & 4.50 & 0.82 & 0.550 & 0.368 & 1 \\
   25329 &  40315 &  351624 &  1732 & -1366 &  -30.0 &  54.14 & 1.08 &  8.51 & 7.18 & 0.58 & 0.870 & 0.529 & 0 \\
   25673 &  40420 &  -43918 &    63 &   185 &   37.0 &  24.23 & 1.53 &  9.56 & 6.48 & 0.70 & 0.816 &       & 0 \\
  284248 &  41436 &  222104 &   426 &  -302 &  339.0 &  12.84 & 1.33 &  9.22 & 4.76 & 0.76 & 0.428 & 0.322 & 0 \\
   29907 &  43822 & -652458 &   733 &  1249 &   56.0 &  17.00 & 0.98 &  9.91 & 6.06 & 0.68 & 0.635 & 0.435 & 1 \\
  280067 &  44030 &  351256 &    53 &   -21 &   -3.0 &  14.69 & 2.05 & 10.40 & 6.24 & 0.71 & 0.696 &       & 0 \\
   29400 &  44250 &  664409 &   355 &    91 &  -59.8 &  27.55 & 1.04 &  8.27 & 5.47 & 0.80 & 0.735 &       & 0 \\
   31128 &  45210 & -270351 &   165 &   -28 &  105.0 &  15.55 & 1.20 &  9.14 & 5.10 & 0.74 & 0.490 & 0.353 & 0 \\
  241253 &  50957 &   53327 &   271 &   -71 &  -16.0 &  10.29 & 1.66 &  9.72 & 4.78 & 0.80 & 0.520 & 0.357 & 0, 0.02 \\
   34328 &  51305 & -593844 &   935 &   515 &  232.0 &  14.55 & 1.01 &  9.46 & 5.27 & 0.72 & 0.484 & 0.365 & 0 \\
   36283 &  53114 &  154624 &   -43 &  -373 &   50.0 &  18.66 & 1.35 &  8.64 & 4.99 & 0.83 & 0.663 & 0.417 & 0 \\
+12~0853 &  54009 &  121042 &   272 &   -70 &   23.2 &  14.30 & 1.99 & 10.22 & 6.00 & 0.70 & 0.650 & 0.435 & 1, 0.02 \\
   40057 &  55726 &  215851 &    15 &   -82 &   39.0 &  16.57 & 1.47 &  9.02 & 5.12 & 0.80 & 0.470 &       & 0 \\
   45205 &  63016 &  604703 &   137 &  -247 &   60.5 &  13.88 & 1.14 &  8.43 & 4.14 & 0.84 & 0.550 & 0.368 & 0 \\
   46341 &  63238 &  -62919 &   256 &     7 &    8.0 &  16.86 & 0.98 &  8.60 & 4.73 & 0.82 & 0.560 & 0.363 & 0 \\
-25~3416 &  63715 & -252140 &   -72 &  -143 &   53.0 &  17.59 & 1.34 &  9.65 & 5.88 & 0.75 & 0.700 &       & 0 \\
-33~3337 &  65448 & -334449 &  -178 &  -149 &   67.0 &   9.11 & 1.01 &  9.03 & 3.83 & 0.82 & 0.471 & 0.337 & 0 \\
   51754 &  65839 &   -2850 &   336 &  -607 &  -94.3 &  14.63 & 1.39 &  9.03 & 4.86 & 0.82 & 0.577 & 0.375 & 0 \\
   53545 &  70449 & -161705 &    22 &   -90 &  -14.0 &  15.74 & 0.91 &  8.06 & 4.05 & 0.88 & 0.468 & 0.299 & 0 \\
-57~1633 &  70629 & -572729 &   -94 &   691 &  282.0 &  10.68 & 0.91 &  9.54 & 4.68 & 0.82 & 0.480 & 0.343 & 1 \\
   53871 &  70927 &  520219 &    33 &    -2 &  -28.0 &  12.91 & 1.45 &  9.07 & 4.62 & 0.85 & 0.455 &       & 0 \\
+17~1524 &  71317 &  172602 &   -55 &  -217 &   -3.7 &  11.23 & 1.78 & 10.27 & 5.52 & 0.80 & 0.800 & 0.475 & 1 \\
   59374 &  73029 &  185741 &    29 &  -437 &   79.9 &  20.00 & 1.66 &  8.50 & 5.01 & 0.79 & 0.520 & 0.367 & 0 \\
-45~3283 &  73419 & -451643 &  -312 &   440 &  317.4 &  15.32 & 1.38 & 10.43 & 6.36 & 0.68 & 0.615 & 0.406 & 0 \\
   60319 &  73435 &  165404 &     5 &  -295 &  -33.7 &  12.15 & 1.24 &  8.95 & 4.37 & 0.84 & 0.510 & 0.350 & 0 \\
   64090 &  75333 &  303618 &   705 & -1835 & -240.0 &  35.29 & 1.04 &  8.30 & 6.04 & 0.67 & 0.618 & 0.430 & 1 \\
   64606 &  75434 &  -12444 &  -252 &   -62 &   93.4 &  52.01 & 1.85 &  7.44 & 6.02 & 0.72 & 0.732 & 0.452 & 1 \\
+23~3511 &  81923 &  540510 &   -34 &  -629 &   59.0 &  10.36 & 1.47 &  9.71 & 4.79 & 0.75 & 0.480 & 0.339 & 0 \\
   74000 &  84051 & -162043 &   352 &  -485 &  204.2 &   7.26 & 1.32 &  9.67 & 3.97 & 0.76 & 0.422 & 0.308 & 0 \\
\hline
\end{tabular}
\normalsize
\label{t:basic}
\end{table*}

\addtocounter{table}{-1}

\begin{table*}
\caption{Basic parameters. Bin=0: bona fide single star; =1: known or suspected binary}
\scriptsize
\begin{tabular}{rrrrrrrrrrrrrl}
\hline
HD/BD &    RA &     dec &$\mu_{\rm RA}$&$\mu_{\rm dec}$&$V_{\rm rad}$&$\pi$&$\Delta\pi$&$V$&$M_V$&Mass&$B-V$&$b-y$&Bin\\
         & HHMMSS &  DDMMSS &  mas  &  mas  &  km/s  &  mas   & mas  & mag & mag & $M\odot$&  mag  &  mag  &   \\       
\hline
   75530 &  85021 &  -53210 &  -182 &  -513 &   33.3 &  18.78 & 1.48 &  9.19 & 5.56 & 0.78 & 0.734 & 0.445 & 0 \\
   76932 &  85843 & -160758 &   244 &   213 &  120.8 &  46.90 & 0.97 &  5.86 & 4.22 & 0.84 & 0.524 & 0.359 & 0 \\
   76910 &  85906 &   -3726 &    77 &    -4 &   35.0 &  12.66 & 1.24 &  8.50 & 4.01 & 0.87 & 0.445 & 0.301 & 0 \\
-03~2525 &  85910 &  -40137 &   345 &  -581 &   25.0 &  12.37 & 1.72 &  9.67 & 5.13 & 0.71 & 0.478 & 0.340 & 1 \\
   78737 &  90903 & -270149 &    70 &   -13 &  -36.0 &   6.48 & 1.10 &  8.95 & 3.01 & 0.87 & 0.410 & 0.293 & 0 \\
-80~0328 &  92421 & -803121 &   202 &  1237 & -110.0 &  16.46 & 0.99 & 10.10 & 6.18 & 0.64 & 0.576 & 0.423 & 0 \\
   83220 &  93517 & -490749 &    27 &    18 &   -9.0 &  10.41 & 0.94 &  8.56 & 3.65 & 0.88 & 0.398 & 0.285 & 1 \\
   83888 &  94247 &  453102 &    -5 &     7 &  -14.0 &  12.30 & 1.16 &  8.84 & 4.29 & 0.89 & 0.424 &       & 0 \\
+09~2242 &  94853 &   85828 &    18 &   -61 &   -2.0 &  13.99 & 1.56 &  9.59 & 5.32 & 0.82 & 0.498 &       & 0 \\
   84937 &  94856 &  134439 &   374 &  -775 &  -16.7 &  12.44 & 1.06 &  8.28 & 3.75 & 0.75 & 0.396 & 0.302 & 1 \\
   88725 & 101433 &   30905 &   230 &  -400 &  -24.2 &  27.67 & 1.01 &  7.74 & 4.95 & 0.82 & 0.601 & 0.397 & 0 \\
   91345 & 103005 & -713339 &  -130 &    52 &   35.0 &  17.70 & 0.93 &  9.04 & 5.28 & 0.76 & 0.558 & 0.387 & 1 \\
+29~2091 & 104723 &  282356 &   179 &  -826 &   74.0 &  10.55 & 1.75 & 10.22 & 5.34 & 0.70 & 0.500 & 0.382 & 0 \\
   94028 & 105128 &  201639 &  -262 &  -456 &   61.9 &  19.23 & 1.13 &  8.23 & 4.65 & 0.77 & 0.474 & 0.344 & 1 \\
   97320 & 111101 & -652538 &   159 &  -201 &   51.1 &  17.77 & 0.76 &  8.15 & 4.40 & 0.80 & 0.482 & 0.337 & 1 \\
   97916 & 111554 &   21512 &   208 &    -8 &   55.4 &   7.69 & 1.23 &  9.17 & 3.60 & 0.84 & 0.420 & 0.292 & 1 \\
  103095 & 115259 &  374307 &  4003 & -5815 &  -98.0 & 109.22 & 0.78 &  6.45 & 6.64 & 0.65 & 0.750 & 0.484 & 0 \\
  105755 & 121016 &  542917 &    64 &   -21 &  -37.0 &  12.80 & 1.01 &  8.59 & 4.13 & 0.85 & 0.568 & 0.385 & 1 \\
  106038 & 121201 &  131541 &  -217 &  -439 &   95.0 &   9.16 & 1.50 & 10.18 & 4.99 & 0.77 & 0.458 & 0.342 & 0 \\
  106516 & 121511 & -101845 &    32 & -1012 &    8.2 &  44.34 & 1.01 &  6.11 & 4.34 & 0.84 & 0.467 & 0.319 & 1 \\
  108076 & 122446 &  381907 &  -587 &    65 &   -1.3 &  26.94 & 0.82 &  8.02 & 5.17 & 0.79 & 0.560 & 0.386 & 1 \\
  108177 & 122535 &   11702 &    34 &  -470 &  159.0 &  10.95 & 1.29 &  9.66 & 4.86 & 0.75 & 0.435 & 0.331 & 1 \\
  111980 & 125315 & -183120 &   300 &  -795 &  144.4 &  12.48 & 1.38 &  8.38 & 3.86 & 0.83 & 0.538 & 0.370 & 0 \\
 113083A & 130126 & -272228 &  -476 &  -202 &  227.3 &  18.51 & 1.12 &  8.05 & 4.39 & 0.83 & 0.540 & 0.367 & 1 \\
 113083B & 130126 & -272228 &  -476 &  -202 &  227.3 &  18.51 & 1.12 &  8.05 & 4.39 & 0.83 & 0.540 & 0.367 & 1 \\
  113679 & 130553 & -383100 &  -388 &  -145 &  156.0 &   6.82 & 1.32 &  9.70 & 3.87 & 0.87 & 0.605 & 0.404 & 0, 0.024 \\
+33~2300 & 130633 &  324001 &    45 &   -29 &  -39.0 &  12.74 & 1.66 & 10.09 & 5.62 & 0.80 & 0.518 &       & 1 \\
  114762 & 131220 &  173101 &  -583 &    -2 &   49.9 &  24.65 & 1.44 &  7.30 & 4.26 & 0.84 & 0.532 & 0.365 & 1 \\
  116064 & 132144 & -391840 &  -751 &   116 &  143.4 &  15.54 & 1.44 &  8.81 & 4.77 & 0.74 & 0.465 & 0.341 & 1, 0.035 \\
  116316 & 132234 &  260657 &  -162 &    17 &  -28.0 &  19.04 & 0.95 &  7.67 & 4.07 & 0.85 & 0.485 & 0.304 & 1 \\
  118659 & 133800 &  190853 &   134 &  -322 &  -45.3 &  18.98 & 1.22 &  8.84 & 5.23 & 0.80 & 0.680 & 0.423 & 0 \\
  119173 & 134143 &  -40146 &  -203 &   -65 &  -88.0 &  17.57 & 1.11 &  8.83 & 5.05 & 0.81 & 0.550 &       & 0 \\
  120559 & 135140 & -572608 &  -360 &  -413 &   13.4 &  40.02 & 1.00 &  7.97 & 5.98 & 0.72 & 0.664 & 0.424 & 0 \\
  121004 & 135358 & -463220 &  -482 &     9 &  243.8 &  16.73 & 1.35 &  9.04 & 5.16 & 0.79 & 0.608 & 0.399 & 1 \\
  123710 & 140457 &  743425 &  -147 &    90 &    6.8 &  22.84 & 0.68 &  8.22 & 5.01 & 0.82 & 0.590 & 0.402 & 0 \\
  126681 & 142725 & -182440 &   -71 &  -313 &  -47.2 &  19.16 & 1.44 &  9.32 & 5.73 & 0.72 & 0.597 & 0.400 & 0 \\
  129515 & 144211 &  315536 &    34 &    52 &   -6.0 &  13.43 & 1.07 &  8.78 & 4.42 & 0.86 & 0.461 &       & 0 \\
  129392 & 144211 &   35619 &   -29 &   -19 &  -16.0 &  11.53 & 1.47 &  8.92 & 4.23 & 0.88 & 0.383 &       & 0 \\
  129518 & 144249 &   40245 &    47 &    46 &  -12.0 &  14.31 & 1.49 &  8.86 & 4.64 & 0.85 & 0.477 &       & 0 \\
+26~2606 & 144902 &  254209 &    -9 &  -347 &   34.0 &  10.28 & 1.42 &  9.72 & 4.78 & 0.70 & 0.424 & 0.332 & 1 \\
  132475 & 145950 & -220046 &  -560 &  -499 &  167.0 &  10.85 & 1.14 &  8.57 & 3.75 & 0.80 & 0.560 & 0.393 & 0, 0.057 \\
  134113 & 150747 &   85247 &  -519 &   -57 &  -60.0 &  15.40 & 1.37 &  8.26 & 4.20 & 0.85 & 0.570 & 0.388 & 1 \\
  134088 & 150813 &  -75448 &  -158 &  -446 &  -59.0 &  28.29 & 1.04 &  8.00 & 5.26 & 0.78 & 0.590 & 0.389 & 0 \\
  134169 & 150818 &   35550 &     1 &   -15 &   18.3 &  16.80 & 1.11 &  7.67 & 3.80 & 0.86 & 0.535 & 0.370 & 1 \\
  134439 & 151013 & -162246 &  -999 & -3542 &  294.3 &  34.14 & 1.36 &  9.07 & 6.74 & 0.64 & 0.777 & 0.486 & 0 \\
  134440 & 151013 & -162747 & -1001 & -3542 &  308.0 &  33.68 & 1.67 &  9.44 & 7.08 & 0.60 & 0.854 & 0.522 & 0 \\
  140283 & 154303 & -105601 & -1116 &  -303 & -171.4 &  17.44 & 0.97 &  7.24 & 3.45 & 0.75 & 0.492 & 0.379 & 0, 0.021 \\
  142575 & 155503 &   50412 &  -281 &    26 &  -64.8 &   6.56 & 1.23 &  8.62 & 2.70 & 0.84 & 0.375 & 0.270 & 0 \\
+42~2667 & 160313 &  421447 &  -195 &  -366 & -157.0 &   8.03 & 1.12 &  9.85 & 4.37 & 0.79 & 0.467 & 0.342 & 0 \\
  145417 & 161349 & -573414 &  -865 & -1403 &   10.0 &  72.75 & 0.82 &  7.52 & 6.83 & 0.64 & 0.820 & 0.505 & 0 \\
\hline
\end{tabular}
\normalsize
\end{table*}

\addtocounter{table}{-1}

\begin{table*}
\caption{Basic parameters. Bin=0: bona fide single star; =1: known or suspected binary}
\scriptsize
\begin{tabular}{rrrrrrrrrrrrrl}
\hline
HD/BD &    RA &     dec &$\mu_{\rm RA}$&$\mu_{\rm dec}$&$V_{\rm rad}$&$\pi$&$\Delta\pi$&$V$&$M_V$&Mass&$B-V$&$b-y$&Bin\\
         & HHMMSS &  DDMMSS &  mas  &  mas  &  km/s  &  mas   & mas  & mag & mag & $M\odot$&  mag  &  mag  &   \\       
\hline
  148816 & 163028 &   41042 &  -433 & -1392 &  -51.7 &  24.34 & 0.90 &  7.27 & 4.20 & 0.84 & 0.538 & 0.367 & 0 \\
  149414 & 163443 &  -41344 &  -133 &  -704 & -169.9 &  20.71 & 1.50 &  9.63 & 6.21 & 0.68 & 0.743 & 0.474 & 1 \\
  149996 & 163817 &  -22632 &  -178 &  -285 &  -35.9 &  14.37 & 1.18 &  8.49 & 4.28 & 0.86 & 0.610 & 0.396 & 0 \\
  157466 & 172228 &  245246 &    63 &  -160 &   28.0 &  33.54 & 0.84 &  6.89 & 4.52 & 0.85 & 0.505 & 0.352 & 0 \\
+31~3025 & 172642 &  310334 &  -359 &    74 &  -73.4 &  14.04 & 1.26 &  9.67 & 5.41 & 0.79 & 0.710 &       & 0 \\
  158226 & 172643 &  310438 &  -362 &    74 &  -70.1 &  14.51 & 0.93 &  8.50 & 4.31 & 0.86 & 0.590 & 0.383 & 0 \\
  158809 & 173125 &  -23220 &  -273 &  -107 &    3.8 &  13.31 & 1.27 &  8.13 & 3.75 & 0.87 & 0.654 & 0.424 & 1 \\
  159482 & 173443 &   60052 &  -479 &   374 & -144.0 &  20.90 & 1.18 &  8.39 & 4.99 & 0.80 & 0.569 & 0.384 & 0 \\
  160693 & 173937 &  371102 &  -498 &  -820 &   39.7 &  18.32 & 0.78 &  8.36 & 4.67 & 0.84 & 0.587 & 0.383 & 1 \\
+02~3375 & 173946 &   22500 &  -366 &    75 & -389.4 &   8.35 & 1.64 &  9.93 & 4.54 & 0.71 & 0.451 & 0.352 & 0 \\
  163810 & 175838 & -130550 &  -456 &  -734 &  191.0 &  11.88 & 2.21 &  9.62 & 4.99 & 0.77 & 0.605 & 0.424 & 1, 0.03 \\
  163799 & 175856 & -222304 &  -287 &  -137 &  -16.0 &  11.14 & 1.50 &  8.81 & 4.04 & 0.84 & 0.540 & 0.367 & 0 \\
+05~3640 & 181222 &   52404 &  -500 &  -646 &   -1.5 &  17.00 & 1.90 & 10.43 & 6.58 & 0.66 & 0.730 & 0.474 & 1 \\
  166913 & 181626 & -592411 &  -254 &  -113 &  -42.6 &  16.09 & 1.04 &  8.23 & 4.26 & 0.78 & 0.450 & 0.327 & 0 \\
  171620 & 183431 &  342456 &   196 &   191 &  -32.3 &  19.12 & 0.69 &  7.56 & 3.97 & 0.87 & 0.500 & 0.340 & 0 \\
  174912 & 185125 &  383736 &   324 &    44 &  -11.9 &  33.31 & 0.61 &  7.16 & 4.77 & 0.84 & 0.540 & 0.371 & 0 \\
  175179 & 185423 &  -43619 &  -133 &  -431 &   17.8 &  11.85 & 1.52 &  9.04 & 4.41 & 0.84 & 0.582 & 0.382 & 0 \\
  179626 & 191321 &   -3542 &  -315 &  -447 &  -70.8 &   7.52 & 1.36 &  9.14 & 3.52 & 0.83 & 0.528 & 0.373 & 0, 0.026 \\
  181743 & 192343 & -450457 &   -65 &  -810 &   18.0 &  11.31 & 1.76 &  9.71 & 4.98 & 0.73 & 0.460 & 0.349 & 0 \\
  184499 & 193327 &  331207 &  -464 &   224 & -163.3 &  31.29 & 0.62 &  6.61 & 4.09 & 0.87 & 0.578 & 0.390 & 0 \\
  186379 & 194307 &  243553 &    87 &  -271 &   -8.3 &  22.10 & 0.82 &  6.76 & 3.48 & 0.88 & 0.555 & 0.374 & 0 \\
  188510 & 195510 &  104427 &   -38 &   290 & -192.8 &  25.32 & 1.17 &  8.82 & 5.84 & 0.69 & 0.587 & 0.416 & 1 \\
  189558 & 200100 & -121520 &  -309 &  -365 &  -14.7 &  14.76 & 1.10 &  7.72 & 3.57 & 0.84 & 0.554 & 0.386 & 0 \\
+42~3607 & 200901 &  425155 &   119 &   341 & -196.4 &  12.04 & 1.13 & 10.11 & 5.51 & 0.68 & 0.510 &       & 0, 0.04 \\
+23~3912 & 201048 &  235755 &  -172 &    63 & -110.5 &   9.38 & 1.24 &  8.93 & 3.79 & 0.81 & 0.510 & 0.372 & 0 \\
  192718 & 201638 &  -72638 &   313 &  -129 & -109.0 &  17.28 & 1.22 &  8.41 & 4.60 & 0.84 & 0.575 & 0.373 & 0 \\
  193901 & 202336 & -212214 &   540 & -1056 & -172.0 &  22.88 & 1.24 &  8.67 & 5.47 & 0.75 & 0.553 & 0.381 & 0 \\
  194598 & 202612 &   92700 &   118 &  -550 & -246.3 &  17.94 & 1.24 &  8.36 & 4.63 & 0.80 & 0.486 & 0.342 & 0 \\
  195633 & 203224 &   63103 &    74 &    24 &  -69.0 &   8.63 & 1.16 &  8.54 & 3.22 & 0.86 & 0.523 & 0.361 & 0, 0.025 \\
  195987 & 203252 &  415354 &  -157 &   453 &  -10.0 &  44.99 & 0.64 &  7.09 & 5.36 & 0.79 & 0.802 & 0.480 & 1 \\
  196892 & 204049 & -184733 &    44 &  -428 &  -30.5 &  15.78 & 1.22 &  8.23 & 4.22 & 0.82 & 0.498 & 0.349 & 0 \\
+41~3931 & 205517 &  421801 &    57 &  -390 & -131.4 &  14.24 & 1.46 & 10.27 & 6.04 & 0.67 & 0.620 &       & 0, 0.03 \\
+33~4117 & 210043 &  335321 &    -1 &  -321 &  -15.4 &  14.86 & 1.42 &  9.62 & 5.48 & 0.80 & 0.750 &       & 0 \\
+19~4601 & 210212 &  195403 &     4 &   223 &  -61.3 &  17.10 & 1.30 &  9.12 & 5.28 & 0.80 & 0.650 & 0.413 & 0 \\
  201891 & 211159 &  174340 &  -122 &  -899 &  -45.1 &  28.26 & 1.01 &  7.38 & 4.64 & 0.80 & 0.510 & 0.358 & 0 \\
  201889 & 211159 &  241005 &   439 &   110 & -102.5 &  17.95 & 1.44 &  8.04 & 4.31 & 0.84 & 0.585 & 0.388 & 0 \\
  204155 & 212643 &   52630 &   167 &  -247 &  -84.6 &  13.02 & 1.11 &  8.47 & 4.04 & 0.86 & 0.570 & 0.378 & 0 \\
  205650 & 213726 & -273807 &   342 &  -208 & -102.1 &  18.61 & 1.23 &  9.00 & 5.35 & 0.75 & 0.530 & 0.374 & 0 \\
+59~2407 & 213916 &  601702 &  -382 &   233 & -245.4 &  15.20 & 1.21 & 10.34 & 6.25 & 0.64 & 0.630 &       & 0, 0.05 \\
+22~4454 & 213936 &  231556 &    32 &   205 & -104.3 &  17.66 & 1.44 &  9.50 & 5.73 & 0.77 & 0.770 & 0.459 & 0 \\
  207978 & 215230 &  284737 &   -59 &   -63 &   19.0 &  36.15 & 0.69 &  5.54 & 3.33 & 0.87 & 0.412 & 0.299 & 0 \\
+11~4725 & 220541 &  122236 &   201 &  -421 & -200.8 &  21.52 & 1.59 &  9.55 & 6.21 & 0.70 & 0.640 & 0.423 & 1 \\
+17~4708 & 221132 &  180534 &   512 &    60 & -295.6 &   8.43 & 1.42 &  9.47 & 4.10 & 0.79 & 0.444 & 0.329 & 1 \\
  211998 & 222437 & -721520 &  1302 &  -674 &   20.5 &  34.60 & 0.60 &  5.28 & 2.98 & 0.82 & 0.658 & 0.448 & 1 \\
  218502 & 230839 & -150312 &   104 &  -286 &  -32.0 &  14.33 & 1.20 &  8.25 & 4.03 & 0.77 & 0.411 & 0.312 & 0 \\
 219175A & 231407 &  -85528 &   551 &   -38 &  -30.2 &  26.52 & 2.41 &  7.56 & 4.68 & 0.84 & 0.558 & 0.358 & 1 \\
 219175B & 231408 &  -85553 &   567 &   -55 &  -29.4 &  35.69 & 5.65 &  8.18 & 5.94 & 0.74 & 0.695 & 0.418 & 1 \\
+33~4707 & 232511 &  341714 &  -298 &  -695 &  -42.0 &  25.02 & 1.34 &  9.35 & 6.34 & 0.74 & 0.967 &       & 0 \\
  221377 & 233120 &  522438 &   101 &   -30 &   27.0 &  11.01 & 0.91 &  7.57 & 2.78 & 0.86 & 0.390 & 0.296 & 1 \\
  222794 & 234327 &  580449 &   390 &   482 &  -67.1 &  22.01 & 0.65 &  7.11 & 3.82 & 0.86 & 0.647 & 0.416 & 0 \\
\hline
\end{tabular}
\normalsize
\end{table*}

Table~\ref{t:basic} contains the most important parameters for the program
stars:
\begin{itemize}
\item Proper motions and parallaxes with their errors are from the Hipparcos
Catalog (Perryman et al. 1997); absolute magnitudes were obtained using 
the apparent $V$\ magnitudes listed in the Hipparcos Catalog, assuming
negligible interstellar absorption 
\item $B-V$ colors are the simple average of all individual $B-V$\ colors
measurements from the Simbad catalog; for a few stars missing this data, it
was obtained from the Hipparcos Catalog
\item $b-y$\ colors are from the catalog by Hauck \& Mermilliod (1998); for
stars from Nissen \& Schuster (1997), they were taken from that paper 
\item radial velocities are from the Simbad catalog, whenever possible; else
they were taken from Barbier-Brossat \& Figon (2000) General Catalog of Mean
Radial Velocities (available through Vizier at CDS). For a few stars missing
these data, they were taken from Carney et al. (1994), Ryan \& Norris (1991),
Norris (1986), or from our own observations
\item the last column give informations about binarity, in most cases derived
from Carney et al. (1994), with few additions from SIMBAD. In this column, we
also listed values of reddening (from Nissen et al. 2002; Carney et al. 1994;
and Schuster \& Nissen 1989). Note that reddening should be negligible for
most program stars
\end{itemize}

\section{Kinematics and stellar populations}

We transformed radial velocities and proper motions into the corresponding
galactocentric velocity components $\Pi, \Theta$, and $Z$, and corrected them
for the Standard Solar Motion and the Motion of the Local Standard of Rest
(LSR). For this last step we have used a solar motion of $(U,V,W) = (+10.0,
+5.2, 7.2 )$~km/s, according to Dehnen \& Binney (1998). The adopted
procedure follows the method of Johnson \& Soderblom (1987); however we
adopted a right-handed reference frame with the x-axis pointed toward the
center. The y-axis is along the the direction of galactic rotation, and
the z-axis is toward the North Galactic Pole.

\subsection{Galactic model of mass distribution}

The equation of motion have been integrated adopting the model for the Galactic
gravitational potential and corresponding mass distribution by Allen \&
Santill\'an (1991).

In this model, the mass distribution of the Galaxy is described as a three
component system: a spherical central bulge, and a flattened disk, both in the
Miyamoto-Nagai form, plus a massive spherical halo. The gravitational
potential is fully analytical, continuous everywhere, and has continuous
derivatives; its simple mathematical form leads to a rapid integration of the
orbits with high numerical precision. The model provides accurate
representation of the Galactic rotation curve $V_C(R)$ and the force $F_z(z)$
perpendicular to the Galactic Plane. The values obtained for the Galactic
rotation constants are $A= 12.95 ~\rm{km}~\rm{s}^{-1}~\rm{kpc}^{-1}$ and
$B=-12.93~\rm{km}~\rm{s}^{-1}~\rm{kpc}^{-1}$, in good agreement with
observational data.

The expression for the potential of the three components are:
\begin{equation}
\phi_B(r,z) = - \frac{G M_B}{\sqrt{r^2+z^2+b_B^2}}
\end{equation}
\begin{equation}
\phi_D(r,z) = - \frac{G M_D}{\sqrt{r^2+\bigg(a_D+\sqrt{z^2+b_D^2}\bigg)^2}}
\end{equation}

\begin{equation}
\phi_H(r,z) = - \frac{G M_H}{\varrho}\cdot \frac{\big(\frac{\varrho}{a_H}\big)^{2.02}}{1+\big(\frac{\varrho}{a_H}\big)^{1.02}}
-
\frac{M_H}{1.02\cdot a_H}\bigg[ -\frac{1.02}{1+\big(\frac{\varrho}{a_H}\big)^{1.02}} + \ln\Big(1+\Big(\frac{\varrho}{a_H}\Big)^
{1.02} \Big) \bigg]^{100}_R
\end{equation}
where $\varrho = \sqrt{r^2+z^2}$. In this expression, $G$\ is the constant of
gravity, $M_B$, $M_D$, $M_H$, $b_B$, $a_D$, $b_D$, and $a_H$\ are the masses
and scale lengths for the Bulge, Disk and Halo respectively. It should also
be noted that while assumptions about these parameters affect the derivation
of the orbital parameters, they are not crucial in our discussion, that is
essentially based on a ranking of the stars according to the different
kinematic parameters.

\noindent
Table \ref{t:tab3} lists the values of the various constants for this model.
The total mass of the model is $9\,10^{11} {\rm M_\odot}$, and the Halo is
truncated at 100 kpc.

\begin{table}
\caption{Constants for the galactic model}
\begin{tabular}{lcrr}
\hline
        &       &                    &\\
galactocentric distance of Sun & $R_\odot$ & 8.5   & kpc \\
local circular velocity        & $\Theta$  & 220   & km ${\rm s}^{-1}$ \\
        &       &                    &\\
Bulge   & $M_B$& 1.41$\cdot 10^{10}$ & ${\rm M}_\odot$ \\
        & $b_B$& 0.3873              & kpc \\
Disk    & $M_D$& 8.56$\cdot 10^{10}$ & ${\rm M}_\odot$ \\
        & $a_D$& 5.3178              & kpc \\
        & $b_D$& 0.2500              & kpc \\
Halo    & $M_H$& 80.02$\cdot 10^{10}$& ${\rm M}_\odot$ \\
        & $a_H$& 12.0                & kpc \\
        &       &                    &\\
\hline
\end{tabular}
\label{t:tab3}
\end{table}

~\begin{table*}
\caption{Kynematics. Pop=0: dissipative component; =1: accretion component;
=2: thin disk}
\scriptsize
\begin{tabular}{rrrrrrrrrr}
\hline
HD/DM    & $U$ & $V$ & $W$ &$R_{\rm min}$&$R_{\rm max}$&$Z_{\rm max}$&$e$&$V_{\rm rot}$&Pop\\
         &   km/s &   km/s &   km/s &   kpc  &  kpc  &  kpc  &       & km/s &   \\  
\hline
  224930 &  $-$8.25 & $-$72.57 & $-$31.51 & 4.579 & 8.503 & 0.310& 0.300 & 152.63& 0 \\
    3567 & 135.80 &$-$235.81 & $-$43.89 & 0.189 &10.721 & 7.616& 0.965 & $-$10.61& 1 \\
    3628 &$-$168.67 & $-$54.81 &  48.91 & 4.279 &13.667 & 1.414& 0.523 & 170.39& 0 \\
$-$35~0360 &$-$110.51 &$-$174.47 & $-$22.23 & 0.958 & 9.500 & 0.382& 0.817 &  50.73& 0 \\
    6582 & $-$43.10 &$-$156.92 & $-$35.31 & 1.474 & 8.617 & 0.439& 0.708 &  68.28& 0 \\
    9430 &  93.45 & $-$20.04 & $-$15.69 & 5.954 &11.803 & 0.131& 0.329 & 205.16& 0 \\
$-$61~0282 & 238.64 &$-$263.82 & $-$43.01 & 0.670 &16.871 & 8.343& 0.924 & $-$38.62& 1 \\
   10607 &   5.71 &$-$148.42 & 114.80 & 2.230 & 8.518 & 4.638& 0.585 &  76.78& 0 \\
+29~0366 & $-$60.02 & $-$77.22 & $-$46.68 & 4.288 & 8.965 & 0.605& 0.353 & 147.98& 0 \\
$-$01~0306 &$-$201.70 &$-$202.51 &  65.71 & 0.401 &12.815 & 6.640& 0.939 &  22.69& 1 \\
   15096 & $-$29.10 & $-$24.59 &  30.91 & 7.185 & 8.725 & 0.558& 0.097 & 200.61& 0 \\
   16397 & 136.52 & $-$31.11 & $-$40.75 & 5.059 &13.802 & 0.655& 0.464 & 194.09& 0 \\
   17288 & $-$99.92 &$-$112.97 &  62.48 & 2.937 & 9.532 & 1.588& 0.529 & 112.22& 0 \\
   17820 &  36.48 &$-$105.14 & $-$81.22 & 3.354 & 8.848 & 1.645& 0.450 & 120.06& 0 \\
   18907 &  37.09 & $-$57.44 &  49.87 & 5.274 & 9.023 & 1.047& 0.262 & 167.76& 0 \\
   19445 & 156.07 &$-$123.05 & $-$67.49 & 2.338 &12.242 & 1.518& 0.679 & 102.15& 0 \\
   20512 &  14.45 & $-$56.93 & $-$57.33 & 5.434 & 8.683 & 0.847& 0.230 & 168.27& 0 \\
$-$47~1087 & $-$88.68 & $-$71.83 &  49.92 & 4.454 & 9.606 & 1.143& 0.366 & 153.37& 0 \\
   22879 &$-$105.99 & $-$84.38 & $-$40.51 & 3.742 & 9.931 & 0.515& 0.453 & 140.82& 0 \\
   23439 & $-$94.50 &$-$115.22 & $-$76.28 & 2.829 & 9.456 & 1.538& 0.539 & 109.98& 0 \\
   24616 & $-$26.25 &$-$162.19 & $-$25.80 & 1.325 & 8.566 & 0.234& 0.732 &  63.01& 0 \\
   25704 &$-$129.77 & $-$62.87 &  $-$6.18 & 4.301 &11.064 & 0.047& 0.440 & 162.33& 0 \\
   25329 & $-$36.89 &$-$190.77 &  21.23 & 0.588 & 8.583 & 4.670& 0.872 &  34.43& 1 \\
   25673 & $-$51.94 &  11.52 &   0.52 & 7.954 &10.972 & 0.106& 0.159 & 236.72& 2 \\
  284248 &$-$352.97 &$-$145.24 & $-$80.03 & 1.406 &36.961 & 1.699& 0.927 &  79.96& 1 \\
   29907 &$-$383.27 &$-$135.31 &  31.89 & 1.506 &47.646 & 2.241& 0.939 &  89.89& 1 \\
  280067 &  $-$1.90 & $-$16.33 &   8.79 & 7.801 & 8.625 & 0.190& 0.050 & 208.87& 2 \\
   29400 &  41.15 & $-$64.13 &  41.80 & 4.908 & 9.048 & 0.840& 0.297 & 161.07& 0 \\
   31128 & $-$59.89 & $-$96.77 & $-$26.03 & 3.476 & 8.869 & 0.239& 0.437 & 128.43& 0 \\
  241253 &  10.59 & $-$93.56 &  89.71 & 4.055 & 8.659 & 2.485& 0.362 & 131.64& 0 \\
   34328 &$-$213.97 &$-$345.35 &  98.64 & 2.653 &15.779 & 4.711& 0.712 &$-$120.15& 1 \\
   36283 & $-$26.38 & $-$80.04 & $-$67.06 & 4.401 & 8.594 & 1.102& 0.323 & 145.16& 0 \\
+12~0853 & $-$17.40 & $-$72.19 &  60.78 & 4.859 & 8.574 & 1.362& 0.277 & 153.01& 0 \\
   40057 & $-$35.63 & $-$27.23 &  $-$8.89 & 6.880 & 8.846 & 0.019& 0.125 & 197.97& 2 \\
   45205 & $-$86.27 & $-$63.69 &  38.30 & 4.707 & 9.710 & 0.790& 0.347 & 161.51& 0 \\
   46341 &   5.27 & $-$33.93 &  63.79 & 6.977 & 8.679 & 1.420& 0.109 & 191.26& 0 \\
$-$25~3416 &  $-$3.54 & $-$51.27 & $-$45.05 & 5.710 & 8.546 & 0.550& 0.199 & 173.93& 0 \\
$-$33~3337 &  12.15 & $-$49.38 &$-$128.43 & 6.672 & 8.877 & 3.434& 0.142 & 175.82& 0 \\
   51754 & 202.39 &$-$135.83 &   4.21 & 1.740 &14.680 & 0.178& 0.788 &  89.37& 0 \\
   53545 &  29.90 &  $-$7.22 &  $-$5.36 & 7.492 & 9.788 & 0.023& 0.133 & 217.98& 2 \\
$-$57~1633 &$-$318.44 &$-$268.46 & $-$43.21 & 0.625 &25.245 & 2.138& 0.952 & $-$43.26& 1 \\
   53871 &  28.30 & $-$11.41 &  $-$0.68 & 7.443 & 9.550 & 0.086& 0.124 & 213.79& 2 \\
+17~1524 &  15.82 & $-$72.28 & $-$58.90 & 4.693 & 8.712 & 0.900& 0.300 & 152.92& 0 \\
   59374 & $-$48.26 &$-$121.35 & $-$10.32 & 2.597 & 8.713 & 0.039& 0.541 & 103.85& 0 \\
$-$45~3283 &$-$227.51 &$-$262.54 & $-$88.96 & 0.661 &14.627 & 1.725& 0.914 & $-$37.34& 1 \\
   60319 &  57.86 & $-$90.86 & $-$52.78 & 3.693 & 9.272 & 0.779& 0.430 & 134.34& 0 \\
   64090 & 265.85 &$-$220.12 & $-$90.56 & 0.177 &20.917 &11.983& 0.983 &   5.08& 1 \\
   64606 & $-$76.62 & $-$58.42 &  $-$0.09 & 4.915 & 9.392 & 0.084& 0.313 & 166.78& 0 \\
+23~3511 &$-$133.41 &$-$261.38 &  21.17 & 0.614 &10.152 & 0.807& 0.886 & $-$36.18& 1 \\
   74000 & 243.24 &$-$362.89 &  62.97 & 2.919 &21.486 & 3.078& 0.761 &$-$137.69& 1 \\
\hline
\end{tabular}
\normalsize
\label{t:kin}
\end{table*}

\addtocounter{table}{-1}

\begin{table*}
\caption{Kynematics. Pop=0: dissipative component; =1: accretion component;
=2: thin disk}
\scriptsize
\begin{tabular}{rrrrrrrrrr}
\hline
HD/DM    & $U$ & $V$ & $W$ &$R_{\rm min}$&$R_{\rm max}$&$Z_{\rm max}$&$e$&$V_{\rm rot}$&Pop\\
         &   km/s &   km/s &   km/s &   kpc  &  kpc  &  kpc  &       & km/s &   \\  
\hline
   75530 &  20.70 &$-$110.79 & $-$85.35 & 3.192 & 8.659 & 1.765& 0.461 & 114.41& 0 \\
   76932 & $-$49.04 & $-$91.18 &  70.32 & 3.950 & 8.754 & 1.738& 0.378 & 134.02& 0 \\
   76910 &  $-$0.96 & $-$25.16 &  37.75 & 7.329 & 8.611 & 0.700& 0.080 & 200.04& 0 \\
$-$03~2525 & 189.29 &$-$178.46 &   2.27 & 0.755 &13.217 & 0.170& 0.892 &  46.74& 0 \\
   78737 &  51.07 &  29.62 &  22.88 & 7.795 &13.587 & 0.555& 0.271 & 254.82& 0 \\
$-$80~0328 &$-$209.31 & $-$87.74 & 301.49 & 3.113 &43.208 &41.822& 0.866 & 137.46& 1 \\
   83220 &   3.06 &   9.75 &  13.96 & 8.369 &10.074 & 0.286& 0.092 & 234.95& 2 \\
   83888 &   7.71 &   1.51 & $-$12.05 & 8.302 & 9.520 & 0.086& 0.068 & 226.71& 2 \\
+09~2242 &  14.16 & $-$15.65 &  $-$4.79 & 7.485 & 8.987 & 0.058& 0.091 & 209.54& 2 \\
   84937 & 226.62 &$-$237.46 &  $-$8.38 & 0.185 &15.584 & 9.082& 0.977 & $-$12.26& 1 \\
   88725 &  72.46 & $-$31.99 & $-$23.69 & 5.795 &10.437 & 0.221& 0.286 & 193.21& 0 \\
   91345 & $-$22.17 & $-$44.45 & $-$12.80 & 5.956 & 8.514 & 0.064& 0.177 & 180.75& 0 \\
+29~2091 & 155.91 &$-$342.91 &  88.35 & 2.887 &12.653 & 1.748& 0.628 &$-$117.71& 1 \\
   94028 & $-$34.19 &$-$139.32 &   7.75 & 2.010 & 8.580 & 0.182& 0.620 &  85.88& 0 \\
   97320 &  73.65 & $-$22.39 & $-$36.87 & 6.103 &10.828 & 0.434& 0.279 & 202.81& 0 \\
   97916 & 106.32 &  10.15 &  90.16 & 6.604 &16.233 & 2.464& 0.422 & 235.35& 1 \\
  103095 & 279.74 &$-$158.32 & $-$13.14 & 1.099 &22.367 & 0.111& 0.906 &  66.88& 1 \\
  105755 &  35.96 &  $-$7.63 & $-$25.30 & 7.397 & 9.943 & 0.250& 0.147 & 217.57& 0 \\
  106038 &  13.36 &$-$269.67 &  18.77 & 0.823 & 8.605 & 0.553& 0.825 & $-$44.47& 1 \\
  106516 &  54.10 & $-$74.38 & $-$57.68 & 4.355 & 9.225 & 0.887& 0.359 & 150.82& 0 \\
  108076 & $-$94.24 & $-$41.49 & $-$14.12 & 5.409 &10.218 & 0.098& 0.308 & 183.71& 0 \\
  108177 & 134.21 &$-$214.17 &  54.95 & 0.206 &10.919 & 6.770& 0.963 &  11.03& 1 \\
  111980 & 267.74 &$-$203.06 &$-$110.06 & 0.326 &21.764 &10.233& 0.970 &  22.14& 1 \\
 113083A &  21.93 &$-$243.44 &  97.25 & 0.377 & 8.633 & 5.193& 0.916 & $-$18.24& 1 \\
 113083B &  21.93 &$-$243.44 &  97.25 & 0.377 & 8.633 & 5.193& 0.916 & $-$18.24& 1 \\
  113679 &$-$118.32 &$-$305.20 &  $-$9.16 & 1.668 & 9.700 & 0.073& 0.706 & $-$80.00& 1 \\
+33~2300 &  20.16 &  $-$3.66 & $-$38.71 & 7.867 & 9.622 & 0.466& 0.100 & 221.54& 2 \\
  114762 & $-$82.85 & $-$69.48 &  58.04 & 4.582 & 9.518 & 1.348& 0.350 & 155.72& 0 \\
  116064 &$-$105.35 &$-$222.59 & 116.86 & 0.077 & 9.653 & 8.883& 0.984 &   2.61& 1 \\
  116316 & $-$37.76 & $-$21.93 & $-$22.84 & 7.097 & 8.903 & 0.199& 0.113 & 203.27& 0 \\
  118659 &  63.02 & $-$43.09 & $-$61.74 & 5.603 & 9.900 & 1.027& 0.277 & 182.10& 0 \\
  119173 & $-$74.19 & $-$19.43 & $-$71.90 & 6.729 &10.352 & 1.363& 0.212 & 205.77& 0 \\
  120559 & $-$28.66 & $-$47.48 & $-$36.27 & 5.815 & 8.572 & 0.381& 0.192 & 177.72& 0 \\
  121004 &  62.28 &$-$252.44 & 102.45 & 0.589 & 9.007 & 6.241& 0.877 & $-$27.24& 1 \\
  123710 & $-$35.59 &  $-$7.71 &   0.14 & 7.794 & 9.250 & 0.091& 0.085 & 217.49& 2 \\
  126681 & $-$21.72 & $-$47.15 & $-$76.41 & 6.140 & 8.504 & 1.345& 0.161 & 178.04& 0 \\
  129515 &  $-$7.48 &  18.46 & $-$10.97 & 8.479 &10.681 & 0.091& 0.115 & 243.66& 2 \\
  129392 & $-$12.59 & $-$13.17 & $-$11.28 & 7.947 & 8.459 & 0.084& 0.031 & 212.03& 2 \\
  129518 &  $-$5.17 &  22.09 & $-$10.19 & 8.452 &11.025 & 0.077& 0.132 & 247.29& 2 \\
+26~2606 & 122.92 &$-$106.09 &  20.36 & 2.834 &10.816 & 0.458& 0.585 & 119.11& 0 \\
  132475 &  34.37 &$-$362.75 &  50.23 & 3.931 & 8.724 & 1.050& 0.379 &$-$137.55& 1 \\
  134113 &$-$114.22 &$-$125.37 &  25.86 & 2.324 & 9.755 & 0.507& 0.615 &  99.83& 0 \\
  134088 & $-$26.83 & $-$67.70 & $-$66.81 & 4.945 & 8.527 & 1.089& 0.266 & 157.50& 0 \\
  134169 &  14.15 &  $-$2.23 &  12.16 & 7.993 & 9.405 & 0.253& 0.081 & 222.97& 0 \\
  134439 & 294.28 &$-$506.03 & $-$71.35 & 5.130 &66.144 & 6.631& 0.856 &$-$280.82& 1 \\
  134440 & 306.20 &$-$514.95 & $-$65.84 & 5.136 &77.716 & 6.626& 0.876 &$-$289.75& 1 \\
  140283 &$-$246.24 &$-$256.77 &  40.22 & 0.507 &15.817 & 2.025& 0.938 & $-$31.57& 1 \\
  142575 &$-$140.06 &$-$131.15 &  94.71 & 2.202 &10.663 & 4.026& 0.658 &  94.05& 0 \\
+42~2667 & 107.85 &$-$268.68 & $-$27.68 & 0.735 & 9.872 & 0.629& 0.861 & $-$43.48& 1 \\
  145417 & $-$47.68 & $-$92.62 & $-$28.00 & 3.648 & 8.686 & 0.259& 0.408 & 132.58& 0 \\
\hline
\end{tabular}
\normalsize
\end{table*}

\addtocounter{table}{-1}

\begin{table*}
\caption{Kynematics. Pop=0: dissipative component; =1: accretion component;
=2: thin disk}
\scriptsize
\begin{tabular}{rrrrrrrrrr}
\hline
HD/DM    & $U$ & $V$ & $W$ &$R_{\rm min}$&$R_{\rm max}$&$Z_{\rm max}$&$e$&$V_{\rm rot}$&Pop\\
         &   km/s &   km/s &   km/s &   kpc  &  kpc  &  kpc  &       & km/s &   \\  
\hline
  148816 &  82.52 &$-$264.71 & $-$80.05 & 0.700 & 9.698 & 5.242& 0.865 & $-$39.51& 1 \\
  149414 & $-$85.88 &$-$172.99 &$-$135.88 & 1.507 & 9.276 & 6.031& 0.721 &  52.21& 0 \\
  149996 &  $-$2.08 &$-$115.18 & $-$17.50 & 2.809 & 8.448 & 0.122& 0.501 & 110.02& 0 \\
  157466 &  35.41 &  10.97 &   0.44 & 7.837 &10.980 & 0.101& 0.167 & 236.17& 2 \\
+31~3025 & $-$66.86 &$-$107.30 &  68.71 & 3.203 & 8.925 & 1.690& 0.472 & 117.90& 0 \\
  158226 & $-$64.29 &$-$103.73 &  67.77 & 3.345 & 8.897 & 1.661& 0.454 & 121.47& 0 \\
  158809 &  15.68 & $-$80.51 &  64.76 & 4.403 & 8.545 & 1.489& 0.320 & 144.69& 0 \\
  159482 &$-$171.00 & $-$65.90 &  78.44 & 3.989 &13.726 & 2.805& 0.550 & 159.30& 0 \\
  160693 & 207.09 &$-$113.05 &  86.82 & 2.501 &16.122 & 1.651& 0.731 & 112.15& 0 \\
+02~3375 &$-$359.23 &$-$247.06 &  80.86 & 0.387 &36.140 &14.205& 0.979 & $-$21.86& 1 \\
  163810 & 273.33 &$-$282.14 &  32.58 & 0.942 &21.207 & 2.366& 0.915 & $-$56.94& 1 \\
  163799 &  $-$2.18 &$-$112.77 &  76.48 & 3.117 & 8.414 & 1.900& 0.459 & 112.43& 0 \\
+05~3640 & 114.49 &$-$192.24 &  42.78 & 0.591 &10.150 & 4.844& 0.890 &  32.96& 1 \\
  166913 & $-$44.31 & $-$44.54 &  67.64 & 6.161 & 8.828 & 1.564& 0.178 & 180.66& 0 \\
  171620 & $-$64.76 &   8.59 & $-$37.15 & 7.572 &11.283 & 0.471& 0.197 & 233.79& 0 \\
  174912 & $-$21.28 &   8.58 & $-$42.20 & 8.406 &10.010 & 0.533& 0.087 & 233.78& 0 \\
  175179 & 103.87 &$-$145.20 & $-$31.70 & 1.726 & 9.826 & 0.405& 0.701 &  80.00& 0 \\
  179626 & 146.50 &$-$315.58 &  52.87 & 1.918 &11.422 & 1.406& 0.712 & $-$90.38& 1 \\
  181743 & $-$48.35 &$-$331.69 & $-$63.06 & 2.704 & 8.684 & 0.947& 0.525 &$-$106.49& 1 \\
  184499 & $-$63.67 &$-$158.87 &  58.87 & 1.455 & 8.824 & 1.255& 0.717 &  66.33& 0 \\
  186379 &  31.93 & $-$27.12 & $-$45.18 & 6.687 & 9.180 & 0.569& 0.157 & 198.08& 0 \\
  188510 &$-$152.56 &$-$113.82 &  62.78 & 2.667 &11.287 & 1.810& 0.618 & 111.38& 0 \\
  189558 &  74.65 &$-$127.84 &  43.49 & 2.342 & 9.305 & 0.999& 0.598 &  97.36& 0 \\
+42~3607 &$-$173.89 &$-$168.39 &  14.15 & 1.013 &11.471 & 0.680& 0.838 &  56.81& 0 \\
+23~3912 & $-$27.67 &$-$100.09 &  99.98 & 3.952 & 8.521 & 2.505& 0.366 & 125.11& 0 \\
  192718 &$-$112.75 & $-$75.81 & $-$45.25 & 3.969 &10.212 & 0.626& 0.440 & 149.38& 0 \\
  193901 &$-$153.29 &$-$246.84 & $-$74.45 & 0.451 &10.587 & 1.871& 0.918 & $-$21.64& 1 \\
  194598 & $-$75.24 &$-$275.94 & $-$30.88 & 0.973 & 8.882 & 0.337& 0.803 & $-$50.74& 1 \\
  195633 & $-$71.99 & $-$37.30 &  $-$3.74 & 5.759 & 9.531 & 0.059& 0.247 & 187.90& 0 \\
  195987 & $-$29.95 &  $-$6.08 &  41.45 & 8.063 & 9.293 & 0.811& 0.071 & 219.12& 0 \\
  196892 &   4.62 &$-$128.06 & $-$34.88 & 2.382 & 8.470 & 0.386& 0.561 &  97.14& 0 \\
+41~3931 &  71.21 &$-$143.37 & $-$94.10 & 1.978 & 9.215 & 2.079& 0.647 &  81.83& 0 \\
+33~4117 &  71.05 & $-$40.50 & $-$63.52 & 5.659 &10.213 & 1.112& 0.287 & 184.70& 0 \\
+19~4601 & $-$63.81 & $-$24.40 &  53.98 & 6.663 & 9.796 & 1.211& 0.190 & 200.80& 0 \\
  201891 &  91.79 &$-$115.59 & $-$58.38 & 2.691 & 9.833 & 1.047& 0.570 & 109.61& 0 \\
  201889 &$-$129.34 & $-$81.62 & $-$37.44 & 3.642 &10.729 & 0.476& 0.493 & 143.58& 0 \\
  204155 & $-$34.27 &$-$125.68 & $-$44.42 & 2.474 & 8.535 & 0.557& 0.551 &  99.52& 0 \\
  205650 &$-$117.00 & $-$84.00 &   8.78 & 3.636 &10.195 & 0.221& 0.474 & 141.20& 0 \\
+59~2407 &  89.81 &$-$244.87 & 108.03 & 0.366 &10.617 & 7.173& 0.933 & $-$19.67& 1 \\
+22~4454 & $-$64.70 & $-$71.29 &  68.65 & 4.696 & 9.101 & 1.691& 0.319 & 153.91& 0 \\
  207978 &  13.59 &  15.83 &  $-$7.32 & 8.308 &10.727 & 0.011& 0.127 & 241.03& 2 \\
+11~4725 & $-$35.62 &$-$220.63 &  30.59 & 0.146 & 8.570 & 6.137& 0.967 &   4.57& 1 \\
+17~4708 &$-$303.94 &$-$281.05 &   6.91 & 0.942 &22.672 & 0.317& 0.920 & $-$55.85& 1 \\
  211998 &$-$147.67 &$-$124.44 & $-$58.97 & 2.321 &10.847 & 1.161& 0.647 & 100.76& 0 \\
  218502 &   2.74 &$-$105.38 &  $-$6.83 & 3.177 & 8.501 & 0.062& 0.456 & 119.82& 0 \\
 219175A & $-$87.62 & $-$53.16 & $-$12.51 & 5.003 & 9.758 & 0.071& 0.322 & 172.04& 0 \\
 219175B & $-$67.14 & $-$45.37 &  $-$4.88 & 5.544 & 9.300 & 0.038& 0.253 & 179.83& 0 \\
+33~4707 & 116.38 & $-$56.40 & $-$74.61 & 4.604 &11.850 & 1.689& 0.440 & 168.80& 0 \\
  221377 & $-$43.07 &   8.16 & $-$29.42 & 8.024 &10.521 & 0.312& 0.135 & 233.36& 0 \\
  222794 & $-$73.53 &$-$100.69 &  82.40 & 3.527 & 9.141 & 2.287& 0.443 & 124.51& 0 \\
\hline
\end{tabular}
\normalsize
\end{table*}

\subsection{Galactic orbits}

The orbits of the stars were computed, and integrated backward in time, over a
time interval equal to $5\,10^9$ years. To perform the numerical
integration, we utilized the Burlish-Stoer method, directly applied to the
second order differential equation that describe the motion of a star. This
numerical method allows to obtain a typical error in energy and in the
z-component of the angular momentum of the star of, respectively, $\Delta E/E
\approx 10^{-4}$ and $\Delta L_z/L_z \approx 10^{-9}$. The main parameters of
the orbits are given in Table~\ref{t:kin}.

\begin{figure*}
\centering
\includegraphics[width=13cm]{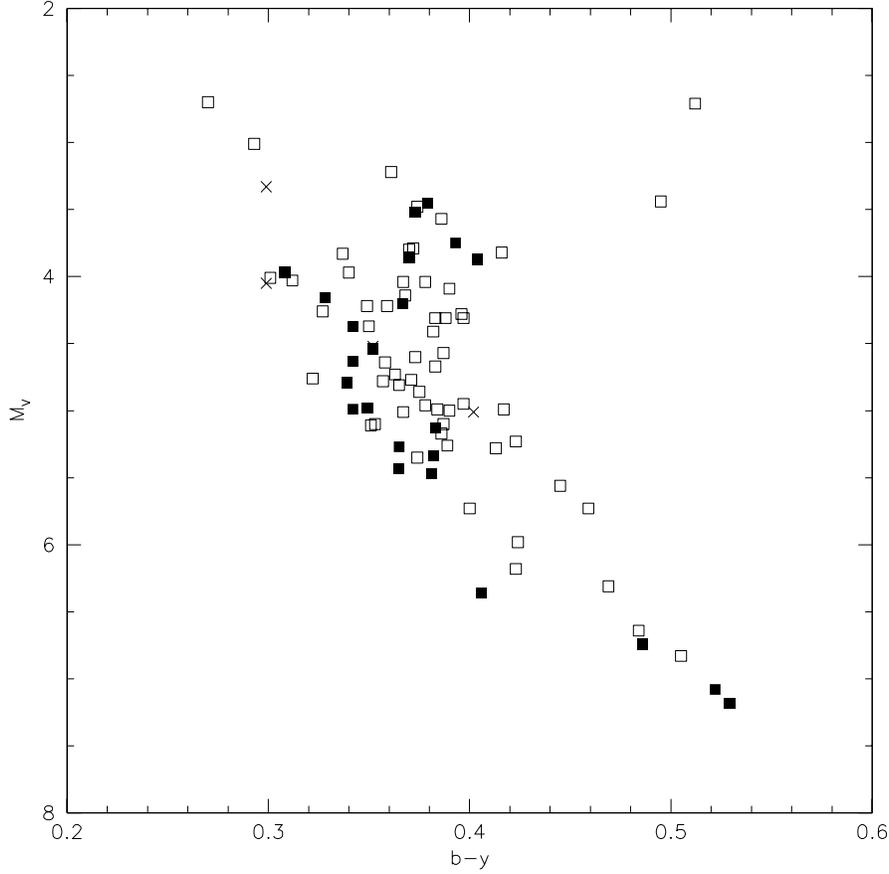}
\caption{ Color magnitude  diagram for the stars considered in this paper
(only bona-fide single stars are plotted). Filled points are accretion
component stars; open points are dissipative component stars; crosses are thin
disk stars  }
\label{f:fcmd}
\end{figure*}  

\subsection{Population membership of the program stars}

In the following, we considered the local stellar sample analyzed in our
program as composed by three distinct populations:
\begin{enumerate}
\item A rotating population, likely produced during the dissipative collapse
of the bulk of the early Galaxy: this population includes part of what usually
is called the Halo as well as the Thick Disk, and can be substantially
identified with the dissipative collapse population of Norris (1994), and with
the population first identified by Eggen et al. (1962). Hereinafter, we will
refer to this population as Dissipative Component. The reason for our choice
is that we are not able to see from our data any clear discontinuity between
the properties of the rotating part of the Halo and those of the Thick Disk
\item A second population composed of non-rotating and counter-rotating stars:
this population includes the remaining part of what is usually called the
Halo, and can be substantially identified with the population due to accretion
processes first proposed by Searle \& Zinn (1978). As we will see in the
next papers, this
population (as a group) has a chemical composition clearly distinct from that
of the first population, likely due to a different origin. We will call these
Accretion Component Stars
\item Finally the Thin Disk, which also has clearly distinct chemical
composition from the first population, as first showed by Gratton et al.
(1996, 2000b), and Fuhrmann (1998). The discontinuity in chemical composition
between the Dissipative Component and the Thin Disk is likely due to a phase
of low star formation that occurred during the early evolution of our Galaxy
(see also Chiappini et al. 1997)
\end{enumerate}

Stars were assigned to the different populations using the following criteria:
\begin{itemize}
\item Thin Disk Stars are those stars for which:
\begin{equation}
\sqrt{Z_{max}^2+4~e^2}<0.35~~{\rm and}~~[{\rm Fe/H}]>-0.7,
\end{equation}
where $Z_{max}$\ is the maximum height above the galactic plane (in kpc), and
$e$\ the eccentricity of the galactic orbit
\item Dissipative Component Stars are those stars that do not belong to the
Thin Disk Stars with a
galactic rotation velocity larger than 40~km/s, an apogalactic
distance $R_{max}>15$~Kpc, and [Fe/H]$<0$
\item All other stars with [Fe/H]$<0$\ are Accretion Component Stars  
\end{itemize}

Of course, these criteria are arbitrary (and depend somewhat on the details of
our dynamical model, for instance on the shape of the adopted galactic
potential); they should then be considered with some care. However, we will
see in the next papers of this series that subdivision of stars in our sample
according to these criteria also corresponds to differences in the chemical
composition, and that with possible caveats about the actual values (but
likely not so much on relative rankings) they likely reflect real differences
in the stellar populations.

According to these definitions, in our sample we have 37 Accretion Component,
99 Dissipative Component, and 14 Thin Disk stars. These numbers certainly do
not reflect the real frequencies of these stars in the solar neighborhood: e.g.
thin disk stars are underrepresented due to the way the sample was
constructed. Fig.~\ref{f:fcmd} shows the location of the program stars in
the color-magnitude (c-m) diagram. There are a few stars whose position seem
quite anomalous: HD97916 and HD142575 are Dissipative Component stars,
but they are bluer and more luminous than the expected turn-off of an old
population; they are likely field blue stragglers (Carney et al. 2001; Glaspey
et al. 1994). CD-45~3283 is much bluer than expected based on luminosity (and
metallicity): we suspect that the parallax is overestimated. HD195633 is too
bright for its color: this star is probably reddened (Schuster \& Nissen
1989).

\begin{table*}
\caption{Atmospheric parameters. $EW$\ source: U=UVES; S=SARG; 
M=McDonald; N=Nissen \& Schuster; F=Fulbright; P=Prochaska}
\scriptsize
\begin{tabular}{rrrrrrrrrr}
\hline
HD/DM    &$T_{\rm eff}$&$T_{\rm eff}$&$T_{\rm eff}$&
          $T_{\rm eff}$&$T_{\rm eff}$&$\log g$&$[$A/H$]$&$v_t$&Source\\
         & Adopted &$B-V$&$b-y$& H$_\alpha$ & IRFM & & & & \\
         &  K  &  K  &  K  &  K  &  K   &      &       & km/s & EW \\
\hline
  224930 & 5357 & 5393 & 5345 &      &      & 4.32 & -0.87 & 0.65 & F  \\
    3567 & 6087 & 6073 & 6136 &      & 5858 & 4.16 & -1.19 & 0.98 & UNF\\
    3628 & 5704 & 5707 & 5666 &      & 5651 & 4.01 & -0.17 & 1.07 & F  \\
-35~0360 & 5048 & 5046 & 5069 & 4957 &      & 4.53 & -1.12 & 0.00 & U  \\
    6582 & 5322 & 5335 & 5304 &      & 5315 & 4.46 & -0.82 & 0.00 & F  \\
    9430 & 5725 & 5716 &      &      &      & 4.36 & -0.34 & 0.68 & F  \\
-61~0282 & 5831 & 5805 & 5857 & 6068 &      & 4.53 & -1.21 & 0.53 & UN \\
   10607 & 5757 & 5733 & 5784 & 5780 &      & 4.01 & -0.94 & 1.50 & U  \\
+29~0366 & 5665 & 5678 & 5658 &      &      & 4.34 & -0.97 & 1.05 & MF \\
-01~0306 & 5726 & 5710 & 5713 &      & 5754 & 4.41 & -0.87 & 0.80 & F  \\
   15096 & 5119 & 5076 &      &      &      & 4.39 & -0.48 & 0.43 & F  \\
   16397 & 5726 & 5755 & 5703 &      &      & 4.21 & -0.50 & 0.85 & N  \\
   17288 & 5744 & 5728 & 5751 &      &      & 4.35 & -0.79 & 0.00 & N  \\
   17820 & 5739 & 5837 & 5622 &      &      & 4.12 & -0.68 & 0.85 & NF \\
   18907 & 5015 & 5026 & 4982 &      &      & 3.49 & -0.83 & 1.41 & F  \\
   19445 & 5976 & 5944 & 5996 &      & 6050 & 4.44 & -2.04 & 1.09 & MF \\
   20512 & 5177 & 5167 & 5121 &      &      & 3.65 & -0.35 & 1.29 & F  \\
-47~1087 & 5672 & 5760 & 5683 &      &      & 4.38 & -0.74 & 0.00 & N  \\
   22879 & 5827 & 5851 & 5850 &      & 5798 & 4.33 & -0.79 & 0.46 & NF \\
   23439 & 5000 & 5055 & 4944 &      &      & 4.49 & -1.04 & 0.03 & MF \\
   24616 & 4943 & 4923 & 4896 &      &      & 3.16 & -0.98 & 1.75 & F  \\
   25704 & 5792 & 5789 & 5821 & 5793 &      & 4.20 & -0.87 & 0.65 & UN \\
   25329 & 4789 &      & 4778 &      & 4842 & 4.68 & -1.76 & 0.00 & F  \\
   25673 & 5008 & 5010 &      &      &      & 4.60 & -0.50 & 0.15 & F  \\
  284248 & 6179 & 6158 & 6212 &      & 6034 & 4.39 & -1.56 & 0.80 & F  \\
   29907 & 5351 & 5393 & 5351 & 5636 &      & 4.57 & -1.48 & 0.00 & UF \\
  280067 & 5480 & 5376 &      &      &      & 4.71 & -0.69 & 0.00 & F  \\
   29400 & 5302 & 5326 &      &      &      & 4.38 & -0.35 & 0.00 & P  \\
   31128 & 5970 & 5894 & 5957 & 6279 &      & 4.45 & -1.50 & 0.89 & UF \\
  241253 & 5897 & 5865 & 5910 &      & 5853 & 4.33 & -1.04 & 0.55 & NP \\
   34328 & 5894 & 5891 & 5872 & 6076 &      & 4.49 & -1.65 & 0.75 & UF \\
   36283 & 5524 & 5564 & 5511 &      &      & 4.30 & -0.34 & 0.60 & P  \\
+12~0853 & 5315 & 5408 & 5321 &      & 5254 & 4.54 & -1.13 & 0.00 & N  \\
   40057 & 6272 & 6243 &      &      &      & 4.59 & -0.52 & 1.15 & F  \\
   45205 & 5805 & 5782 & 5819 &      &      & 4.07 & -0.87 & 1.04 & F  \\
   46341 & 5831 & 5839 & 5874 &      &      & 4.30 & -0.60 & 0.79 & F  \\
-25~3416 & 5391 & 5383 &      &      &      & 4.55 & -0.56 & 0.35 & F  \\
-33~3337 & 6079 & 5987 & 6057 &      &      & 4.03 & -1.25 & 0.70 & N  \\
   51754 & 5807 & 5794 & 5788 &      &      & 4.34 & -0.56 & 0.74 & F  \\
   53545 & 6373 & 6322 & 6428 &      &      & 4.23 & -0.25 & 1.04 & F  \\
-57~1633 & 6013 & 6082 & 6030 &      &      & 4.34 & -0.80 & 0.80 & N  \\
   53871 & 6401 & 6373 &      &      &      & 4.45 & -0.31 & 1.30 & F  \\
+17~1524 & 5079 & 5094 & 5093 &      &      & 4.30 & -0.41 & 0.30 & P  \\
   59374 & 5896 & 5922 & 5841 &      &      & 4.42 & -0.84 & 1.00 & F  \\
-45~3283 & 5692 & 5625 &      &      &      & 4.82 & -0.81 & 0.00 & N  \\
   60319 & 5962 & 5972 & 5969 &      &      & 4.21 & -0.75 & 0.82 & NF \\
   64090 & 5429 & 5429 & 5397 &      & 5441 & 4.59 & -1.59 & 0.31 & MF \\
   64606 & 5218 & 5198 & 5189 &      & 5134 & 4.52 & -0.90 & 0.32 & F  \\
+23~3511 & 5977 & 5896 & 6063 &      &      & 4.34 & -1.66 & 1.05 & F  \\
   74000 & 6216 & 6106 & 6313 &      & 6224 & 4.09 & -1.96 & 1.15 & F  \\
\hline
\end{tabular}
\normalsize
\label{t:par}
\end{table*}

\addtocounter{table}{-1}

\begin{table*}
\caption{Atmospheric parameters. $EW$\ source: U=UVES; S=SARG; 
M=McDonald; N=Nissen \& Schuster; F=Fulbright; P=Prochaska}
\scriptsize
\begin{tabular}{rrrrrrrrrr}
\hline
HD/DM    &$T_{\rm eff}$&$T_{\rm eff}$&$T_{\rm eff}$&
         $T_{\rm eff}$&$T_{\rm eff}$&$\log g$&$[$A/H$]$&$v_t$&Source\\
         & Adopted &$B-V$&$b-y$& H$_\alpha$ & IRFM & & & & \\
         &  K  &  K  &  K  &  K  &  K   &      &       & km/s & EW \\
\hline
   75530 & 5255 & 5264 & 5266 &      &      & 4.39 & -0.57 & 0.00 & P  \\
   76932 & 5923 & 5901 & 5895 &      & 5727 & 4.14 & -0.80 & 1.00 & NF \\
   76910 & 6379 & 6326 & 6389 &      &      & 4.21 & -0.50 & 1.45 & F  \\
-03~2525 & 5850 & 5847 &      &      & 5809 & 4.41 & -2.05 & 0.67 & F  \\
   78737 & 6434 & 6427 & 6418 &      &      & 3.83 & -0.56 & 1.51 & F  \\
-80~0328 & 5496 & 5498 & 5478 & 5495 &      & 4.65 & -1.99 & 1.00 & U  \\
   83220 & 6503 & 6570 & 6523 &      &      & 4.11 & -0.39 & 1.10 & N  \\
   83888 & 6660 & 6644 &      &      &      & 4.42 &  0.01 & 1.20 & F  \\
+09~2242 & 6349 & 6366 &      &      &      & 4.70 &  0.08 & 0.90 & F  \\
   84937 & 6290 & 6223 & 6366 &      & 6330 & 4.02 & -2.18 & 1.25 & SMF\\
   88725 & 5641 & 5689 & 5617 &      & 5669 & 4.32 & -0.61 & 0.60 & F  \\
   91345 & 5663 & 5726 & 5681 & 5648 &      & 4.43 & -1.05 & 0.90 & U  \\
+29~2091 & 5784 & 5780 & 5749 &      &      & 4.46 & -1.89 & 0.59 & F  \\
   94028 & 5995 & 5956 & 6016 &      & 6001 & 4.30 & -1.44 & 1.26 & MF \\
   97320 & 6044 & 5959 & 6068 & 5996 &      & 4.23 & -1.25 & 1.30 & U  \\
   97916 & 6412 & 6316 & 6435 &      & 6393 & 4.05 & -0.87 & 1.26 & F  \\
  103095 & 5025 & 5068 & 4982 &      & 5029 & 4.63 & -1.28 & 0.02 & MF \\
  105755 & 5753 & 5763 & 5704 &      & 5806 & 4.05 & -0.71 & 1.04 & F  \\
  106038 & 6013 & 6078 & 6041 &      & 5960 & 4.44 & -1.23 & 0.00 & N  \\
  106516 & 6232 & 6178 & 6235 &      & 6208 & 4.29 & -0.68 & 1.02 & NF \\
  108076 & 5761 & 5803 & 5697 &      &      & 4.43 & -0.74 & 0.62 & F  \\
  108177 & 6133 & 6111 & 6142 & 5884 & 6097 & 4.41 & -1.65 & 0.90 & UF \\
  111980 & 5745 & 5728 & 5789 & 5723 & 5624 & 3.93 & -1.19 & 1.81 & UF \\
 113083A & 5752 & 5805 & 5823 &      &      & 4.14 & -0.91 & 0.90 & N  \\
 113083B & 5641 & 5803 & 5815 &      &      & 4.10 & -0.88 & 0.80 & N  \\
  113679 & 5543 & 5635 & 5566 &      &      & 3.88 & -0.67 & 0.60 & N  \\
+33~2300 & 6245 & 6298 &      &      &      & 4.78 &  0.11 & 0.60 & F  \\
  114762 & 5875 & 5890 & 5852 &      & 5884 & 4.14 & -0.73 & 1.20 & F  \\
  116064 & 5964 & 5933 & 6051 & 5770 &      & 4.32 & -1.81 & 0.00 & U  \\
  116316 & 6258 & 6105 & 6347 &      &      & 4.19 & -0.67 & 1.40 & F  \\
  118659 & 5446 & 5433 & 5427 &      &      & 4.35 & -0.60 & 0.60 & F  \\
  119173 & 5934 & 5898 &      &      &      & 4.46 & -0.57 & 0.85 & F  \\
  120559 & 5383 & 5418 & 5395 & 5207 &      & 4.57 & -0.90 & 0.00 & UN \\
  121004 & 5583 & 5634 & 5592 & 5401 &      & 4.37 & -0.73 & 0.00 & UN \\
  123710 & 5712 & 5770 & 5599 &      & 5768 & 4.37 & -0.48 & 0.70 & F  \\
  126681 & 5574 & 5593 & 5584 & 5444 & 5541 & 4.55 & -1.10 & 0.14 & UNF\\
  129515 & 6311 & 6348 &      &      &      & 4.35 & -0.27 & 0.80 & F  \\
  129392 & 6698 & 6757 &      &      &      & 4.40 & -0.19 & 1.35 & F  \\
  129518 & 6331 & 6280 &      &      &      & 4.44 & -0.30 & 1.30 & F  \\
+26~2606 & 6101 & 6080 & 6160 &      & 6037 & 4.34 & -2.40 & 0.83 & F  \\
  132475 & 5541 & 5543 & 5607 & 5596 & 5788 & 3.79 & -1.62 & 1.27 & UF \\
  134113 & 5718 & 5751 & 5680 &      &      & 4.06 & -0.72 & 0.95 & F  \\
  134088 & 5674 & 5686 & 5667 &      &      & 4.44 & -0.79 & 0.90 & F  \\
  134169 & 5850 & 5849 & 5808 & 6106 & 5870 & 3.95 & -0.79 & 1.10 & UF \\
  134439 & 4996 & 4977 & 4972 & 5151 & 4974 & 4.65 & -1.33 & 0.01 & UMF\\
  134440 & 4714 & 4712 & 4742 & 4777 & 4746 & 4.61 & -1.42 & 0.00 & UF \\
  140283 & 5657 & 5662 & 5723 & 5560 & 5691 & 3.69 & -2.57 & 1.42 & UF \\
  142575 & 6517 & 6474 & 6571 &      &      & 3.72 & -1.00 & 1.40 & F  \\
+42~2667 & 6021 & 5977 & 6027 &      & 6059 & 4.20 & -1.44 & 1.10 & F  \\
  145417 & 4869 & 4836 & 4842 & 5096 &      & 4.62 & -1.33 & 0.90 & U  \\
\hline
\end{tabular}
\normalsize
\end{table*}

\addtocounter{table}{-1}

\begin{table*}
\caption{Atmospheric parameters. $EW$\ source: U=UVES; S=SARG; 
M=McDonald; N=Nissen \& Schuster; F=Fulbright; P=Prochaska}
\scriptsize
\begin{tabular}{rrrrrrrrrr}
\hline
HD/DM    &$T_{\rm eff}$&$T_{\rm eff}$&$T_{\rm eff}$&
          $T_{\rm eff}$&$T_{\rm eff}$&$\log g$&$[$A/H$]$&$v_t$&Source\\
         & Adopted &$B-V$&$b-y$& H$_\alpha$ & IRFM & & & & \\
         &  K  &  K  &  K  &  K  &  K   &      &       & km/s & EW \\
\hline
  148816 & 5857 & 5863 & 5835 &      & 5851 & 4.11 & -0.74 & 1.10 & F  \\
  149414 & 5080 & 5080 & 5061 &      & 5364 & 4.51 & -1.34 & 0.00 & F  \\
  149996 & 5660 & 5679 & 5637 &      &      & 4.08 & -0.52 & 0.95 & F  \\
  157466 & 6052 & 6124 & 5995 &      &      & 4.31 & -0.36 & 1.05 & F  \\
+31~3025 & 5360 & 5374 &      &      & 5334 & 4.38 & -0.46 & 0.00 & P  \\
  158226 & 5738 & 5753 & 5732 &      & 5794 & 4.12 & -0.51 & 0.95 & F  \\
  158809 & 5478 & 5462 & 5428 &      & 5418 & 3.80 & -0.74 & 1.23 & F  \\
  159482 & 5713 & 5749 & 5706 & 5546 &      & 4.35 & -0.80 & 0.96 & U  \\
  160693 & 5748 & 5772 & 5732 &      & 5771 & 4.26 & -0.50 & 0.85 & F  \\
+02~3375 & 5917 & 5922 & 5972 &      & 5891 & 4.19 & -2.32 & 0.50 & UF \\
  163810 & 5422 & 5497 & 5408 &      & 5493 & 4.23 & -1.29 & 0.90 & F  \\
  163799 & 5827 & 5822 & 5827 &      &      & 4.04 & -0.84 & 1.20 & F  \\
+05~3640 & 5023 &      & 5034 & 4980 & 4980 & 4.61 & -1.15 & 0.00 & U  \\
  166913 & 6070 & 6038 & 6146 & 6100 &      & 4.17 & -1.50 & 1.10 & U  \\
  171620 & 6103 & 6102 & 6072 &      & 6109 & 4.11 & -0.45 & 1.30 & F  \\
  174912 & 5916 & 5972 & 5839 &      &      & 4.36 & -0.42 & 0.85 & F  \\
  175179 & 5745 & 5736 & 5727 &      &      & 4.16 & -0.66 & 0.92 & F  \\
  179626 & 5737 & 5753 & 5762 &      & 5597 & 3.79 & -1.18 & 1.20 & F  \\
  181743 & 5968 & 5969 & 5995 &      &      & 4.40 & -1.77 & 0.72 & U  \\
  184499 & 5764 & 5790 & 5686 &      & 5750 & 4.05 & -0.51 & 1.00 & F  \\
  186379 & 5876 & 5918 & 5827 &      &      & 3.85 & -0.35 & 1.10 & F  \\
  188510 & 5503 & 5556 & 5483 & 5412 & 5564 & 4.55 & -1.42 & 0.51 & UF \\
  189558 & 5668 & 5673 & 5670 & 5829 & 5663 & 3.79 & -1.14 & 1.36 & UF \\
+42~3607 & 5965 & 5742 &      &      &      & 4.58 & -2.06 & 0.52 & S  \\
+23~3912 & 5766 & 5773 & 5773 &      & 5788 & 3.90 & -1.42 & 1.10 & MF \\
  192718 & 5780 & 5780 & 5796 &      &      & 4.24 & -0.61 & 1.05 & F  \\
  193901 & 5779 & 5759 & 5734 & 5848 & 5750 & 4.54 & -1.05 & 1.08 & UF \\
  194598 & 6023 & 5978 & 6033 & 6064 & 6017 & 4.31 & -1.12 & 1.26 & UMF\\
  195633 & 5936 & 5916 & 5878 &      &      & 3.76 & -0.68 & 1.28 & F  \\
  195987 & 5052 & 4983 & 5014 &      &      & 4.22 & -0.83 & 0.60 & F  \\
  196892 & 5893 & 5907 & 5958 & 5928 &      & 4.12 & -1.12 & 1.10 & U  \\
+41~3931 & 5410 & 5401 &      &      &      & 4.58 & -1.71 & 0.95 & F  \\
+33~4117 & 5261 & 5289 &      &      &      & 4.37 & -0.31 & 0.00 & P  \\
+19~4601 & 5528 & 5558 & 5508 &      &      & 4.40 & -0.51 & 0.00 & P  \\
  201891 & 5917 & 5889 & 5902 &      & 5909 & 4.28 & -1.08 & 1.27 & MF \\
  201889 & 5662 & 5682 & 5673 &      & 5635 & 4.08 & -0.79 & 1.05 & F  \\
  204155 & 5772 & 5762 & 5754 & 5907 & 5771 & 4.03 & -0.69 & 0.98 & UF \\
  205650 & 5810 & 5822 & 5786 & 5628 &      & 4.50 & -1.11 & 0.80 & U  \\
+59~2407 & 5391 & 5333 &      &      &      & 4.63 & -1.97 & 0.26 & S  \\
+22~4454 & 5170 & 5154 & 5168 &      & 5267 & 4.41 & -0.55 & 0.43 & F  \\
  207978 & 6418 & 6434 & 6385 &      &      & 3.96 & -0.55 & 1.50 & F  \\
+11~4725 & 5461 & 5525 & 5403 &      & 5411 & 4.69 & -0.80 & 0.22 & F  \\
+17~4708 & 6085 & 6046 & 6129 &      & 5941 & 4.12 & -1.58 & 0.79 & UF \\
  211998 & 5211 & 5241 & 5240 & 5168 &      & 3.36 & -1.53 & 1.20 & U  \\
  218502 & 6242 & 6182 & 6283 &      &      & 4.13 & -1.83 & 0.90 & F  \\
 219175A & 5756 & 5844 &      & 5856 &      & 4.26 & -0.58 & 0.80 & U  \\
 219175B & 5337 & 5390 &      & 5437 &      & 4.55 & -0.58 & 0.00 & U  \\
+33~4707 & 4520 & 4525 &      &      &      & 4.29 & -0.54 & 0.30 & F  \\
  221377 & 6418 & 6459 & 6376 &      &      & 3.73 & -0.73 & 1.50 & F  \\
  222794 & 5473 & 5486 & 5479 &      &      & 3.83 & -0.73 & 0.95 & F  \\
\hline
\end{tabular}
\normalsize
\end{table*}

\begin{figure}
\centering
\includegraphics[width=13cm]{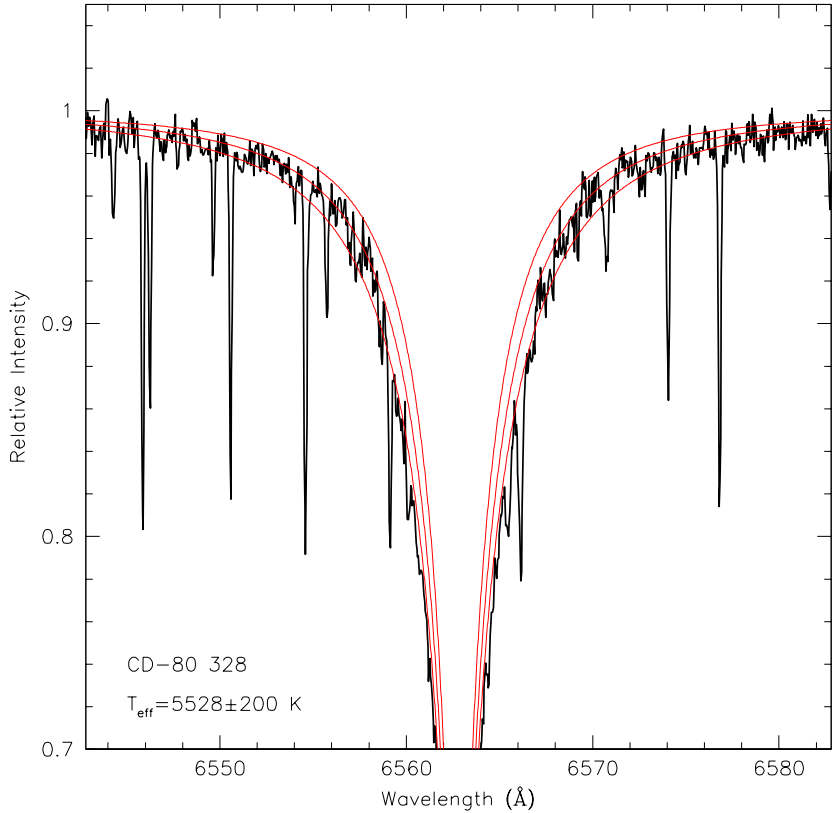}
\caption{ Derivation of temperature from the H$_\alpha$\ profile for one star
of the Large Program (CD$-80~328$). The thick line represents the observed 
profile; the thin lines represent expected profiles for $T_{\rm eff}$=5328,
5528, and 5728~K. Region used in the fitting are those where the relative
intensity is $>$0.9 }
\label{f:halfa}
\end{figure}  

\begin{figure}
\centering
\includegraphics[width=8.8cm]{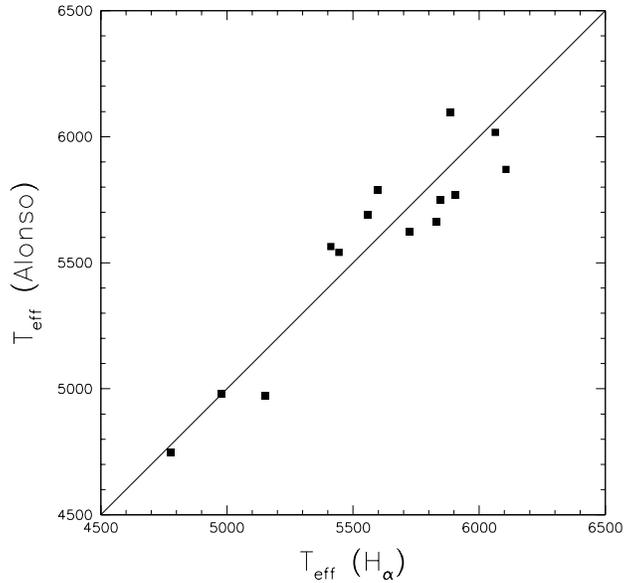}
\caption{ Comparison between temperatures derived by fitting H$_\alpha$\
profiles from spectra of the Large Program, and those obtained by Alonso et
al. (1996) by application of the Infrared Flux Method }
\label{f:talo}
\end{figure}  

\begin{figure}
\centering
\includegraphics[width=8.8cm]{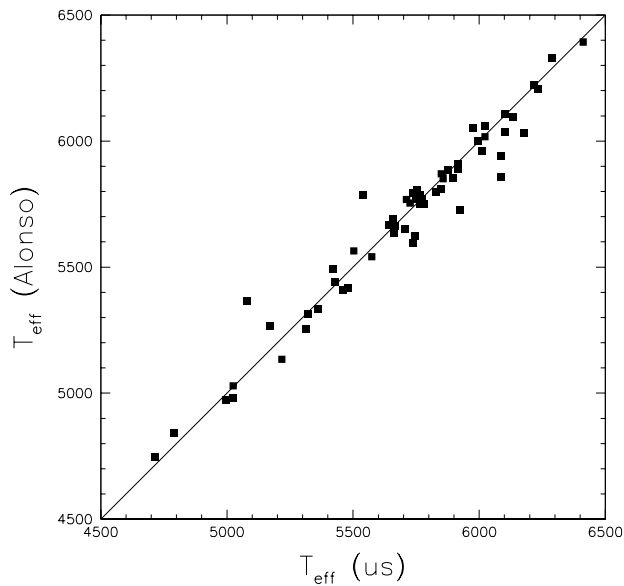}
\caption{ Comparison between temperatures derived from our calibration of
$B-V$\ and $b-y$\ colors, and those obtained by Alonso et al. (1996)
by application of the Infrared Flux Method }
\label{f:talo2}
\end{figure}  

\begin{figure}
\centering
\includegraphics[width=8.8cm]{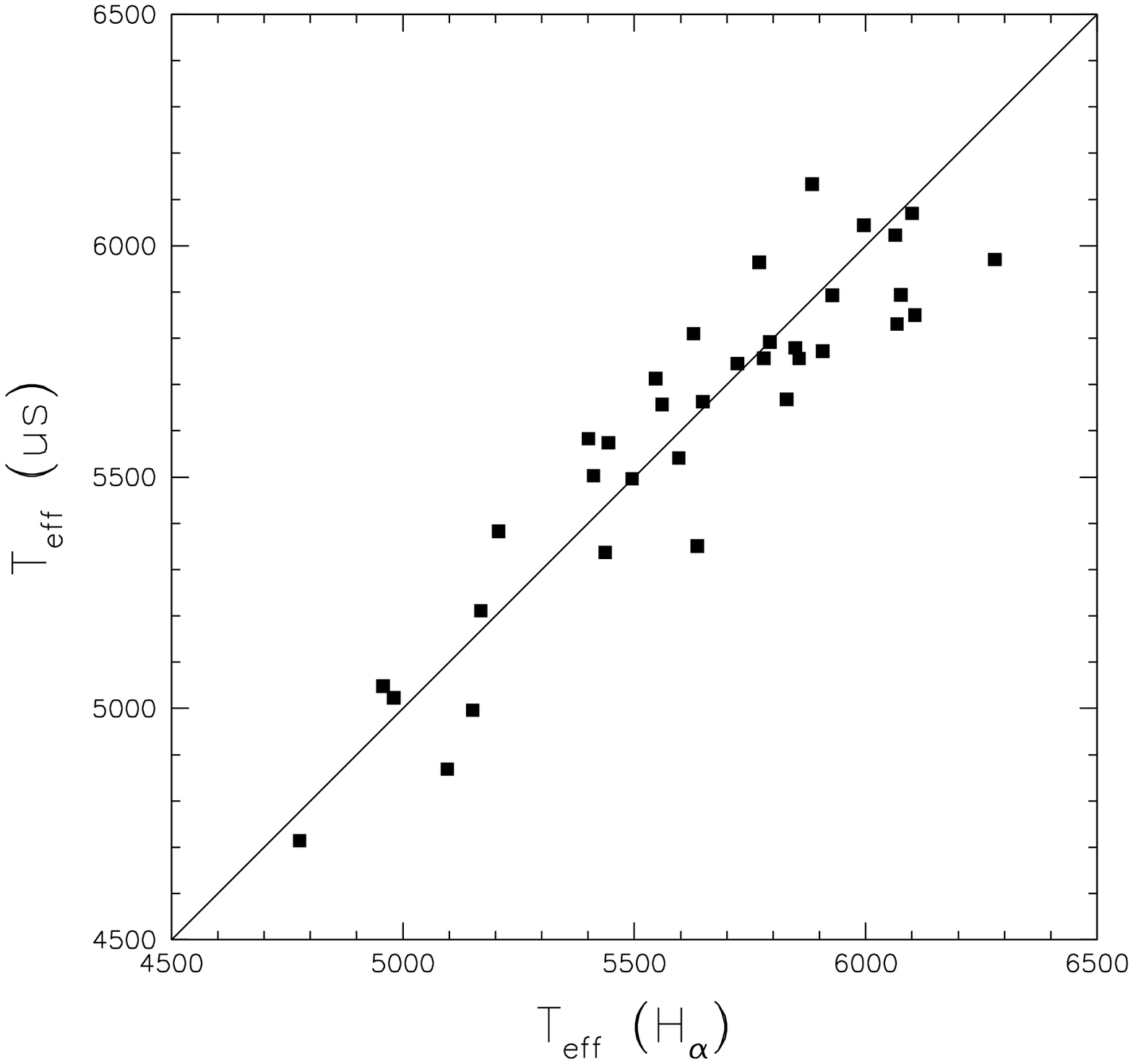}
\caption{ Comparison between temperatures derived from our calibration of
$B-V$\ and $b-y$\ colors, and those obtained by fitting H$_\alpha$\ profiles }
\label{f:tour}
\end{figure}  

\section{Atmospheric Parameters}

Model atmospheres for the program stars were extracted by interpolation within
the grid by Kurucz (1994). Model atmospheres used throughout this paper were
computed with the overshooting option switched off (see discussion in
Castelli et al. 1997).

\subsection{Effective Temperatures}

The atmospheric parameters adopted throughout this paper are listed in
Table~\ref{t:par}. They are consistent with those used in Paper I. Briefly,
whenever possible we did not use directly effective temperatures $T_{\rm
eff}$\ derived from colors, because it is not clear whether the reddening scales
used for globular clusters and field stars are indeed consistent with each
other. The zero point for our $T_{\rm eff}$'s for both globular clusters and
field stars were instead obtained from an analysis of the H$_\alpha$\ profiles
(see Fig.~\ref{f:halfa} for an example of these temperature derivations).
Fig.~\ref{f:talo} compares effective temperatures from H$_\alpha$\ profiles
with those given by Alonso et al. (1996) by application of the Infrared Flux
Method. The agreement is good: on average, our temperatures are higher by
$29\pm 36$~K, with an r.m.s of 140~K (based on 15 stars). We will thereafter
assume that our $T_{\rm eff}$'s coincide with those of Alonso et al. (1996).
However, since $T_{\rm eff}$'s from H$_\alpha$\ profiles have rather large
errors ($\sim 140$~K) for individual stars, due to uncertainties in the flat
fielding and fitting procedure, they were averaged with those given by $B-V$\
and $b-y$\ colors (that have much lower random errors for a given star; we
assumed no reddening for all field stars, since they are all very close to the
Sun). Note that the temperatures from colours listed in Cols.~3 and 4 of
Table~\ref{t:par}
are slightly different from those used in computing the adopted temperatures
listed in Col.~2 of this Table,
because of slightly different assumptions about gravity and metal abundance
when performing the iterative process used in the derivation of temperatures
from colours.
We used the Alonso et al. $T_{\rm eff}$'s (available for 58 stars) to
correct $T_{\rm eff}$'s derived from colors (using the Kurucz calibration) to
the scale given by H$_\alpha$. The result of this calibration is shown in
Fig.~\ref{f:talo2}, that compares our assumed $T_{\rm eff}$'s  with those of
Alonso et al.. The average difference is now $8\pm 11$~K, with an r.m.s.
scatter of the residuals of 83 K; once a few discrepant cases are eliminated
with a 2.5~$\sigma$\ clipping procedure (HD 3567, HD76932, HD132475, and
HD149414), the r.m.s of the residuals for the remaining 54 stars is 54~K. Most
of this remaining scatter can be still attributed to a few other stars
(HD284248, HD111980, HD179626, HD201891, BD+22~4454, and BD+17~4708):
excluding also these stars the r.m.s. scatter is only 38~K. We note that one
of the most discrepant stars (HD 149414) is a known binary, and two others
(HD132475 and HD179626) are likely reddened (see Carney et al. 1994, and
Nissen et al. 2002). We conclude that typical errors in temperatures for
individual stars are about $\sim 50$~K, although there are a few stars for
which errors may be much larger.

As a final comparison, we plotted in Fig.~\ref{f:tour} temperatures derived
from our color calibration against the temperatures derived from the
H$_\alpha$\ profiles. No systematic difference is obvious from this figure.

\subsubsection{Surface gravities}

Surface gravities $g$\ were obtained from the basic relation:
\begin{equation}
\log g = 4\,\log (T_{\rm eff}/5770)+0.4~\,(M_V+BC-4.72)+\log M+4.44
\end{equation}
where $M_V$\ is the absolute visual magnitude (obtained from the apparent
magnitude and the parallax), $BC$\ is the bolometric correction (from Kurucz
CD-ROM15), and $M$\ is the stellar mass. This last was obtained interpolating
the position of the star along isochrones with ages of 14 Gyr from the Padua
group (Girardi et al. 2002). Errors in these surface gravities are mainly due
to errors in the absolute magnitudes, which on turn may be attributed to
uncertainties in the parallaxes; typical values are about $\pm 0.1$~dex.

\subsubsection{Microturbulent velocities and metal abundances}

Microturbulent velocities $v_t$\ were obtained by eliminating trends in the
abundances derived from individual Fe~I lines from expected equivalent widths.
Given the quality of our $EW$s, uncertainties are approximately 
$2/\sqrt(n)$~km/s,
where $n$\ is the number of lines used. $v_t$'s estimated in this way are
well correlated with stellar luminosities. Neglecting variations in stellar
masses, we found that the relation:
\begin{equation}
v_t = 1.127\,[4\,\log (T_{\rm eff}/5770)-\log g+4.44] +0.61~~~~~{\rm km/s}
\end{equation}
is able to predict $v_t$'s values with an accuracy of 0.37~km/s. This is the
error expected when the number of lines used to estimate $v_t$\ is $n=30$.
Hence, for those cases where less than 30 Fe~I lines were measured, we used in
our analysis $v_t$'s predicted by Eq. (6). We conclude that typical errors for
$v_t$'s used throughout this paper are $\pm 0.3$~km/s.

Model metal abundances were set equal to the average Fe abundances derived
from neutral lines. Since uncertainties in Fe abundances are $\sim 0.05$~dex,
we may adopt this value as an estimate of the uncertainty in this parameter.

\begin{table*}
\caption{Abundances}
\scriptsize
\begin{tabular}{rrrrrrrrrrrr}
\hline
HD/DM & $[$Fe/H$]$ & $[$Fe/H$]$ & $[$O/Fe$]$& $[$O/Fe$]$ & $[$Na/Fe$]$ & $[$Mg/Fe$]$ & $[$Si/Fe$]$ & $[$Ca/Fe$]$ &
$[$Ti/Fe$]$ & $[$Ti/Fe$]$ & $[\alpha$/Fe$]$ \\
         &   I   &  II-I &$[$OI$]$&  OI   &   I   &   I   &   I   &   I   &   I   &   II  &       \\
\hline
  224930 & -0.90 & -0.00 &       &       &  0.23 &  0.42 &  0.36 &  0.37 &  0.32 &  0.24 &  0.36 \\
    3567 & -1.22 & -0.04 &  0.59 &  0.46 & -0.28 &  0.22 &  0.17 &  0.32 &  0.17 &  0.27 &  0.23 \\
    3628 & -0.21 & -0.04 &       &       &  0.11 &  0.37 &  0.12 &  0.13 &  0.07 &  0.09 &  0.18 \\
-35~0360 & -1.15 & -0.17 &       &  0.85 & -0.04 &  0.61 &  0.33 &  0.32 &  0.31 &  0.29 &  0.39 \\
    6582 & -0.87 & -0.06 &       &       &  0.08 &  0.33 &  0.34 &  0.28 &  0.33 &  0.25 &  0.31 \\
    9430 & -0.37 & -0.13 &       &       & -0.00 &  0.35 &  0.13 &  0.15 &  0.24 &  0.15 &  0.21 \\
-61~0282 & -1.25 &  0.04 &       &  0.44 & -0.28 &  0.23 &  0.19 &  0.31 &  0.19 &  0.28 &  0.24 \\
   10607 & -0.99 & -0.16 &       &  0.71 &  0.00 &  0.62 &  0.32 &  0.30 &  0.29 &  0.25 &  0.38 \\
+29~0366 & -1.02 & -0.12 &  0.49 &  0.73 &  0.07 &  0.48 &  0.23 &  0.22 &  0.21 &  0.14 &  0.28 \\
-01~0306 & -0.91 & -0.12 &       &       & -0.05 &  0.48 &  0.26 &  0.29 &  0.22 &  0.16 &  0.30 \\
   15096 & -0.53 &  0.30 &       &       &  0.12 &  0.37 &  0.28 &  0.15 &  0.07 &  0.25 &  0.24 \\
   16397 & -0.54 & -0.07 &       &       &  0.07 &  0.32 &  0.17 &  0.16 &  0.11 &  0.09 &  0.19 \\
   17288 & -0.82 & -0.11 &       &  0.49 & -0.05 &  0.37 &  0.14 &  0.26 &  0.27 &  0.25 &  0.26 \\
   17820 & -0.72 & -0.07 &       &  0.52 &  0.11 &  0.43 &  0.27 &  0.28 &  0.21 &  0.19 &  0.29 \\
   18907 & -0.88 &  0.00 &       &       &  0.24 &  0.55 &  0.44 &  0.33 &  0.23 &  0.24 &  0.39 \\
   19445 & -2.08 & -0.05 &       &  0.81 &  0.07 &  0.52 &  0.48 &  0.34 &  0.29 &  0.27 &  0.40 \\
   20512 & -0.40 &  0.07 &       &       &  0.12 &  0.25 &  0.19 &  0.13 & -0.03 & -0.01 &  0.14 \\
-47~1087 & -0.79 &  0.04 &       &  0.49 & -0.01 &  0.42 &  0.25 &  0.26 &  0.27 &  0.42 &  0.32 \\
   22879 & -0.83 & -0.06 &       &  0.53 &  0.03 &  0.45 &  0.24 &  0.28 &  0.19 &  0.25 &  0.30 \\
   23439 & -1.07 &  0.01 &  0.56 &  0.92 & -0.07 &  0.52 &  0.44 &  0.30 &  0.25 &  0.28 &  0.38 \\
   24616 & -1.04 &  0.02 &       &       &  0.26 &  0.63 &  0.51 &  0.35 &  0.15 &  0.25 &  0.42 \\
   25704 & -0.91 & -0.10 &  0.23 &  0.56 &  0.14 &  0.31 &  0.19 &  0.25 &  0.13 &  0.10 &  0.22 \\
   25329 & -1.80 & -0.06 &       &       &  0.24 &  0.59 &  0.51 &  0.51 &  0.40 &  0.30 &  0.49 \\
   25673 & -0.53 &  0.16 &       &       &  0.13 &  0.15 &  0.33 &  0.05 &  0.05 &       &  0.14 \\
  284248 & -1.60 & -0.03 &       &       &  0.12 &  0.35 &       &  0.33 &  0.29 &  0.27 &  0.32 \\
   29907 & -1.52 & -0.01 &       &  0.71 & -0.23 &  0.31 &  0.18 &  0.29 &  0.18 &  0.25 &  0.25 \\
  280067 & -0.66 &  0.29 &       &       &  0.09 &  0.56 &  0.43 &  0.18 &  0.23 &  0.47 &  0.38 \\
   29400 & -0.39 &  0.08 &       &  0.41 &  0.02 &  0.42 &  0.22 &  0.21 &  0.30 &  0.34 &  0.29 \\
   31128 & -1.54 & -0.04 &       &  0.71 & -0.14 &  0.45 &  0.33 &  0.40 &  0.39 &  0.49 &  0.41 \\
  241253 & -1.08 & -0.11 &       &  0.59 & -0.02 &  0.42 &  0.26 &  0.27 &  0.18 &  0.21 &  0.29 \\
   34328 & -1.69 & -0.08 &       &  0.69 & -0.10 &  0.45 &  0.38 &  0.40 &  0.26 &  0.29 &  0.37 \\
   36283 & -0.40 & -0.03 &       &  0.38 &  0.12 &  0.44 &  0.21 &  0.26 &  0.24 &  0.26 &  0.29 \\
+12~0853 & -1.17 &  0.02 &       &  0.59 &  0.00 &  0.39 &  0.27 &  0.25 &  0.34 &  0.22 &  0.30 \\
   40057 & -0.57 & -0.12 &       &       &  0.00 &  0.12 &  0.04 &  0.05 &  0.02 & -0.01 &  0.05 \\
   45205 & -0.91 &  0.05 &       &       &  0.10 &  0.35 &  0.32 &  0.25 &  0.18 &  0.18 &  0.28 \\
   46341 & -0.64 &  0.05 &       &       &  0.04 &  0.24 &  0.20 &  0.13 &       & -0.05 &  0.13 \\
-25~3416 & -0.60 &  0.04 &       &       &  0.09 &  0.53 &  0.32 &  0.20 &  0.28 &  0.27 &  0.33 \\
-33~3337 & -1.28 & -0.07 &       &  0.49 &  0.06 &  0.44 &  0.18 &  0.25 &  0.26 &  0.26 &  0.28 \\
   51754 & -0.60 &       &       &       &  0.01 &  0.51 &  0.23 &  0.22 &       &  0.29 &  0.31 \\
   53545 & -0.29 &  0.03 &       &       & -0.02 &  0.14 &  0.08 &  0.12 &  0.08 &       &  0.10 \\
-57~1633 & -0.84 & -0.07 &       &  0.17 & -0.38 &  0.11 & -0.01 &  0.13 &  0.04 &  0.04 &  0.07 \\
   53871 & -0.35 & -0.16 &       &       &  0.05 &  0.12 &  0.03 &  0.08 &  0.06 &  0.01 &  0.07 \\
+17~1524 & -0.44 &  0.05 &       &  0.37 &  0.14 &  0.37 &  0.16 &  0.15 &  0.21 &  0.23 &  0.23 \\
   59374 & -0.88 & -0.06 &       &       &  0.06 &  0.43 &  0.25 &  0.25 &  0.21 &  0.25 &  0.29 \\
-45~3283 & -0.85 &  0.07 &       &  0.34 & -0.39 &  0.16 &  0.08 &  0.11 &  0.13 &  0.21 &  0.13 \\
   60319 & -0.80 & -0.05 &       &  0.36 &  0.07 &  0.33 &  0.17 &  0.21 &  0.20 &  0.26 &  0.24 \\
   64090 & -1.64 & -0.22 &       &  0.64 & -0.20 &  0.31 &  0.07 &  0.32 &  0.20 &  0.18 &  0.22 \\
   64606 & -0.94 &  0.04 &       &       &  0.15 &  0.51 &  0.39 &  0.28 &  0.31 &  0.27 &  0.37 \\
+23~3511 & -1.70 &  0.06 &       &       & -0.10 &  0.46 &  0.38 &  0.43 &  0.31 &  0.29 &  0.39 \\
   74000 & -2.02 &       &       &       &  0.64 &  0.38 &       &  0.48 &       &  0.08 &  0.31 \\
\hline
\end{tabular}
\label{t:abund0}
\normalsize
\end{table*}

\addtocounter{table}{-1}

\begin{table*}
\caption{Abundances}
\scriptsize
\begin{tabular}{rrrrrrrrrrrr}
\hline
HD/DM & $[$Fe/H$]$ & $[$Fe/H$]$ & $[$O/Fe$]$& $[$O/Fe$]$ & $[$Na/Fe$]$ & $[$Mg/Fe$]$ & $[$Si/Fe$]$ & $[$Ca/Fe$]$ &
$[$Ti/Fe$]$ & $[$Ti/Fe$]$ & $[\alpha$/Fe$]$ \\
         &   I   &   II  &$[$OI$]$&  OI   &   I   &   I   &   I   &   I   &   I   &   II  &       \\
\hline
   75530 & -0.61 & -0.02 &       &  0.62 &  0.18 &  0.46 &  0.23 &  0.24 &  0.35 &  0.29 &  0.31 \\
   76932 & -0.85 & -0.10 &  0.55 &  0.47 &  0.04 &  0.41 &  0.23 &  0.28 &  0.24 &  0.23 &  0.29 \\
   76910 & -0.57 & -0.04 &       &       &  0.08 &  0.26 &  0.18 &  0.16 &  0.01 &  0.05 &  0.16 \\
-03~2525 & -2.10 &  0.31 &       &       &       &  0.49 &  0.50 &  0.43 &  0.33 &  0.57 &  0.47 \\
   78737 & -0.60 & -0.02 &       &       &  0.20 &  0.18 &  0.27 &  0.12 &       &       &  0.19 \\
-80~0328 & -2.03 & -0.22 &       &  0.78 & -0.30 &  0.22 &       &  0.13 &  0.07 &  0.25 &  0.17 \\
   83220 & -0.43 & -0.06 &       &  0.07 &  0.04 &  0.11 &  0.02 &  0.09 & -0.06 &  0.08 &  0.06 \\
   83888 & -0.02 & -0.05 &       &       & -0.07 &  0.03 & -0.04 &  0.04 & -0.14 & -0.04 & -0.02 \\
+09~2242 &  0.04 & -0.06 &       &       & -0.11 & -0.01 & -0.08 &  0.00 &  0.01 &  0.15 &  0.00 \\
   84937 & -2.22 &  0.06 &       &       & -0.41 &  0.52 &  0.59 &  0.44 &  0.41 &  0.41 &  0.49 \\
   88725 & -0.66 & -0.04 &       &       &  0.07 &  0.35 &  0.22 &  0.22 &  0.19 &  0.14 &  0.24 \\
   91345 & -1.09 &  0.02 &       &  0.72 & -0.18 &  0.35 &  0.35 &  0.36 &  0.33 &  0.39 &  0.36 \\
+29~2091 & -1.93 &       &       &       & -0.02 &  0.45 &       &  0.44 &       &  0.39 &  0.43 \\
   94028 & -1.49 & -0.07 &       &  0.68 & -0.07 &  0.53 &  0.35 &  0.32 &  0.30 &  0.23 &  0.36 \\
   97320 & -1.28 &  0.03 &  0.41 &  0.38 &  0.09 &  0.43 &  0.23 &  0.27 &  0.25 &  0.30 &  0.30 \\
   97916 & -0.91 & -0.06 &       &       &  0.19 &  0.53 &  0.40 &  0.33 &       &  0.31 &  0.39 \\
  103095 & -1.33 & -0.07 &       &  0.62 & -0.36 &  0.43 &  0.19 &  0.29 &  0.26 &  0.25 &  0.29 \\
  105755 & -0.75 &  0.01 &       &       &  0.11 &  0.40 &  0.28 &  0.20 &       &  0.18 &  0.26 \\
  106038 & -1.27 & -0.06 &       &  0.58 &  0.07 &  0.47 &  0.60 &  0.25 &  0.18 &  0.28 &  0.39 \\
  106516 & -0.72 & -0.09 &       &  0.46 &  0.12 &  0.47 &  0.31 &  0.29 &  0.14 &  0.15 &  0.30 \\
  108076 & -0.80 & -0.06 &       &       &  0.04 &  0.39 &  0.25 &  0.22 &  0.20 &  0.13 &  0.26 \\
  108177 & -1.69 &  0.00 &       &  0.77 & -0.12 &  0.46 &  0.47 &  0.36 &  0.28 &  0.48 &  0.41 \\
  111980 & -1.23 & -0.02 &  0.43 &       &  0.06 &  0.60 &  0.42 &  0.35 &  0.20 &  0.22 &  0.40 \\
 113083A & -0.94 & -0.12 &       &  0.36 & -0.08 &  0.26 &  0.07 &  0.14 &  0.02 & -0.03 &  0.14 \\
 113083B & -0.92 & -0.08 &       &  0.43 & -0.16 &  0.26 &  0.08 &  0.11 & -0.06 &  0.04 &  0.16 \\
  113679 & -0.70 &  0.04 &  0.19 &  0.55 &  0.05 &  0.41 &  0.27 &  0.31 &  0.21 &  0.42 &  0.33 \\
+33~2300 &  0.07 & -0.00 &       &       & -0.18 & -0.04 & -0.11 & -0.04 & -0.03 &  0.09 & -0.04 \\
  114762 & -0.79 & -0.06 &       &       &  0.16 &  0.47 &  0.26 &  0.23 &  0.16 &  0.17 &  0.28 \\
  116064 & -1.84 & -0.15 &       &  0.60 & -0.09 &  0.40 &  0.24 &  0.33 &  0.27 &  0.38 &  0.33 \\
  116316 & -0.71 &  0.05 &       &       &  0.19 &  0.25 &  0.23 &  0.22 &  0.02 &  0.09 &  0.19 \\
  118659 & -0.65 & -0.03 &       &       &  0.05 &  0.33 &  0.19 &  0.21 &  0.19 &  0.12 &  0.22 \\
  119173 & -0.62 & -0.06 &       &       &  0.01 &  0.26 &  0.13 &  0.13 &  0.08 &  0.06 &  0.15 \\
  120559 & -0.94 &  0.05 &       &  0.58 &  0.06 &  0.46 &  0.31 &  0.24 &  0.27 &  0.33 &  0.33 \\
  121004 & -0.76 &  0.05 &  0.27 &  0.53 &  0.07 &  0.49 &  0.30 &  0.27 &  0.25 &  0.30 &  0.33 \\
  123710 & -0.52 & -0.04 &       &       &  0.06 &  0.16 &  0.08 &  0.10 &  0.04 &  0.02 &  0.09 \\
  126681 & -1.14 & -0.10 &  0.56 &  0.60 & -0.14 &  0.39 &  0.36 &  0.32 &  0.27 &  0.17 &  0.32 \\
  129515 & -0.31 & -0.15 &       &       & -0.06 &  0.09 &  0.02 &  0.08 & -0.04 & -0.05 &  0.03 \\
  129392 & -0.22 & -0.09 &       &       &  0.03 &  0.12 &  0.04 &  0.06 & -0.04 &  0.00 &  0.05 \\
  129518 & -0.34 & -0.07 &       &       &  0.06 &  0.18 &  0.08 &  0.09 & -0.02 &  0.03 &  0.09 \\
+26~2606 & -2.47 &  0.04 &       &       &  0.19 &  0.47 &       &  0.51 &  0.52 &  0.59 &  0.51 \\
  132475 & -1.67 &  0.02 &  0.47 &  0.82 &  0.10 &  0.56 &  0.57 &  0.40 &  0.17 &  0.32 &  0.44 \\
  134113 & -0.76 & -0.08 &       &       &  0.12 &  0.49 &  0.27 &  0.25 &  0.18 &  0.24 &  0.30 \\
  134088 & -0.83 & -0.01 &       &       &  0.11 &  0.45 &  0.32 &  0.25 &  0.23 &  0.26 &  0.32 \\
  134169 & -0.84 & -0.04 &       &  0.54 &  0.08 &  0.44 &  0.23 &  0.26 &  0.20 &  0.18 &  0.28 \\
  134439 & -1.38 & -0.14 &       &  0.36 & -0.56 &  0.19 &  0.01 &  0.14 &  0.11 &  0.14 &  0.12 \\
  134440 & -1.45 &  0.15 &       &       & -0.58 &  0.27 &  0.24 &  0.15 &  0.05 &  0.15 &  0.19 \\
  140283 & -2.61 &  0.08 &  0.87 &  0.80 & -0.15 &  0.40 &  0.32 &  0.36 &  0.38 &  0.38 &  0.37 \\
  142575 & -1.04 &  0.10 &       &       &  0.25 &  0.61 &  0.41 &  0.42 &  0.23 &  0.36 &  0.43 \\
+42~2667 & -1.48 & -0.02 &       &       & -0.09 &  0.47 &  0.38 &  0.42 &  0.28 &  0.29 &  0.39 \\
  145417 & -1.39 & -0.18 &       &  0.81 &  0.00 &       &  0.34 &  0.37 &  0.37 &  0.20 &  0.34 \\
\hline
\end{tabular}
\normalsize
\end{table*}

\addtocounter{table}{-1}

\begin{table*}
\caption{Abundances}
\scriptsize
\begin{tabular}{rrrrrrrrrrrr}
\hline
HD/DM & $[$Fe/H$]$ & $[$Fe/H$]$ & $[$O/Fe$]$& $[$O/Fe$]$ & $[$Na/Fe$]$ & $[$Mg/Fe$]$ & $[$Si/Fe$]$ & $[$Ca/Fe$]$ &
$[$Ti/Fe$]$ & $[$Ti/Fe$]$ & $[\alpha$/Fe$]$ \\
         &   I   &   II  &$[$OI$]$&  OI   &   I   &   I   &   I   &   I   &   I   &   II  &       \\
\hline
  148816 & -0.79 & -0.09 &       &       &  0.17 &  0.49 &  0.29 &  0.26 &  0.21 &  0.19 &  0.31 \\
  149414 & -1.38 &  0.04 &       &       &  0.07 &  0.62 &  0.46 &  0.36 &  0.35 &  0.31 &  0.44 \\
  149996 & -0.57 & -0.09 &       &       &  0.16 &  0.49 &  0.29 &  0.26 &  0.25 &  0.18 &  0.31 \\
  157466 & -0.41 & -0.06 &       &       &  0.01 &  0.20 &  0.04 &  0.05 & -0.06 & -0.08 &  0.05 \\
+31~3025 & -0.49 & -0.05 &       &  0.44 &  0.04 &  0.37 &  0.20 &  0.26 &  0.37 &  0.32 &  0.29 \\
  158226 & -0.55 & -0.08 &       &       &  0.12 &  0.45 &  0.26 &  0.28 &  0.24 &  0.24 &  0.31 \\
  158809 & -0.78 & -0.12 &       &       &  0.21 &  0.60 &  0.34 &  0.38 &  0.25 &  0.16 &  0.38 \\
  159482 & -0.84 & -0.00 &       &  0.45 &  0.18 &  0.52 &  0.34 &  0.31 &  0.22 &  0.33 &  0.36 \\
  160693 & -0.54 & -0.04 &       &       &  0.04 &  0.33 &  0.18 &  0.17 &  0.12 &  0.15 &  0.20 \\
+02~3375 & -2.37 &  0.04 &       &  0.74 &       &  0.49 &       &  0.39 &  0.40 &  0.44 &  0.43 \\
  163810 & -1.36 & -0.08 &       &       & -0.18 &  0.41 &  0.17 &  0.36 &  0.17 &  0.08 &  0.27 \\
  163799 & -0.88 & -0.10 &       &       &  0.08 &  0.49 &  0.37 &  0.34 &  0.23 &  0.25 &  0.36 \\
+05~3640 & -1.19 &  0.08 &       &  0.78 &  0.11 &  0.66 &  0.63 &  0.32 &  0.23 &  0.24 &  0.46 \\
  166913 & -1.54 & -0.18 &  0.59 &  0.74 & -0.08 &  0.47 &  0.34 &  0.37 &  0.28 &  0.35 &  0.37 \\
  171620 & -0.51 & -0.09 &       &       &  0.12 &  0.28 &  0.17 &  0.14 &  0.04 &  0.07 &  0.16 \\
  174912 & -0.46 & -0.11 &       &       &  0.04 &  0.22 &  0.08 &  0.10 &  0.01 &  0.01 &  0.10 \\
  175179 & -0.71 & -0.11 &       &       &  0.13 &  0.44 &  0.26 &  0.30 &  0.21 &  0.10 &  0.29 \\
  179626 & -1.22 & -0.04 &       &       &  0.14 &  0.53 &  0.36 &  0.40 &  0.22 &  0.18 &  0.37 \\
  181743 & -1.81 & -0.10 &       &  0.63 &  0.05 &  0.49 &  0.38 &  0.29 &  0.26 &  0.36 &  0.37 \\
  184499 & -0.55 & -0.13 &       &       &  0.12 &  0.49 &  0.25 &  0.23 &  0.21 &  0.19 &  0.29 \\
  186379 & -0.39 & -0.07 &       &       &  0.05 &  0.26 &  0.07 &  0.14 & -0.00 &  0.02 &  0.12 \\
  188510 & -1.45 & -0.19 &       &  0.79 & -0.37 &  0.19 &  0.21 &  0.25 &  0.20 &  0.13 &  0.20 \\
  189558 & -1.18 & -0.05 &  0.50 &  0.52 & -0.13 &  0.51 &  0.36 &  0.35 &  0.21 &  0.26 &  0.36 \\
+42~3607 & -2.10 &       &       &       &       &  0.49 &       &  0.26 &  0.29 &  0.11 &  0.32 \\
+23~3912 & -1.46 & -0.03 &       &  0.88 & -0.07 &  0.49 &  0.34 &  0.39 &  0.23 &  0.37 &  0.38 \\
  192718 & -0.65 & -0.03 &       &       &  0.15 &  0.45 &  0.28 &  0.26 &  0.21 &  0.23 &  0.30 \\
  193901 & -1.09 & -0.07 &       &  0.40 & -0.31 &  0.20 &  0.06 &  0.21 &  0.11 &  0.13 &  0.15 \\
  194598 & -1.17 & -0.06 &  0.46 &  0.46 & -0.05 &  0.33 &  0.17 &  0.26 &  0.14 &  0.17 &  0.23 \\
  195633 & -0.73 &  0.00 &  0.13 &       &  0.18 &  0.36 &  0.22 &  0.20 &  0.04 &  0.06 &  0.21 \\
  195987 & -0.88 &  0.21 &       &       &  0.38 &  0.37 &  0.56 &  0.32 &  0.31 &  0.35 &  0.39 \\
  196892 & -1.16 & -0.08 &  0.73 &  0.84 &  0.15 &  0.52 &  0.41 &  0.35 &  0.25 &  0.31 &  0.39 \\
+41~3931 & -1.75 & -0.08 &       &       & -0.37 &  0.37 &  0.23 &  0.29 &  0.18 &  0.12 &  0.26 \\
+33~4117 & -0.34 &  0.02 &       &  0.25 & -0.09 &  0.30 &  0.06 &  0.14 &  0.17 &  0.12 &  0.16 \\
+19~4601 & -0.55 &  0.00 &       &  0.35 &  0.06 &  0.42 &  0.20 &  0.21 &  0.25 &  0.31 &  0.28 \\
  201891 & -1.10 & -0.13 &  0.76 &  0.56 &  0.09 &  0.46 &  0.26 &  0.26 &  0.19 &  0.18 &  0.29 \\
  201889 & -0.83 & -0.07 &       &       &  0.19 &  0.53 &  0.34 &  0.42 &  0.25 &  0.21 &  0.38 \\
  204155 & -0.73 & -0.07 &       &       &  0.17 &  0.52 &  0.29 &  0.30 &  0.24 &  0.25 &  0.34 \\
  205650 & -1.16 & -0.05 &  0.53 &  0.59 &  0.08 &  0.45 &  0.32 &  0.30 &  0.17 &  0.21 &  0.32 \\
+59~2407 & -2.01 & -0.03 &       &       &  0.14 &  0.51 &  0.39 &  0.31 &  0.34 &       &  0.39 \\
+22~4454 & -0.60 &  0.03 &       &       &  0.20 &  0.44 &  0.28 &  0.24 &  0.29 &  0.17 &  0.30 \\
  207978 & -0.59 & -0.07 &       &       &  0.17 &  0.31 &  0.15 &  0.17 &  0.07 &  0.03 &  0.17 \\
+11~4725 & -0.84 & -0.02 &       &       &  0.02 &  0.45 &  0.31 &  0.28 &  0.29 &  0.25 &  0.32 \\
+17~4708 & -1.62 &  0.03 &       &  0.70 & -0.01 &  0.47 &  0.28 &  0.42 &  0.36 &  0.30 &  0.37 \\
  211998 & -1.56 & -0.06 &  0.42 &  0.67 &  0.10 &  0.52 &  0.41 &  0.32 &  0.20 &  0.26 &  0.37 \\
  218502 & -1.87 &  0.04 &       &       &  0.17 &  0.45 &       &  0.47 &  0.38 &  0.35 &  0.43 \\
 219175A & -0.63 &  0.10 & -0.03 &  0.23 &  0.07 &  0.28 &  0.21 &  0.05 & -0.10 &  0.09 &  0.13 \\
 219175B & -0.63 &  0.25 &  0.05 &  0.19 &  0.12 &  0.25 &  0.20 &  0.05 & -0.10 &  0.23 &  0.14 \\
+33~4707 & -0.58 &  0.42 &       &       &  0.07 &  0.41 &  0.54 &  0.05 & -0.12 &  0.25 &  0.27 \\
  221377 & -0.77 & -0.10 &       &       &  0.24 &  0.38 &  0.13 &  0.20 &  0.04 &  0.08 &  0.19 \\
  222794 & -0.79 & -0.05 &       &       &  0.16 &  0.54 &  0.32 &  0.29 &  0.22 &  0.28 &  0.35 \\
\hline
\end{tabular}
\normalsize
\end{table*}

\begin{table*}
\caption{Abundances}
\scriptsize
\begin{tabular}{rrrrrrrr}
\hline
HD/DM & $[$Sc/Fe$]$ & $[$V/Fe$]$ & $[$Cr/Fe$]$ & $[$Cr/Fe$]$ & $[$Mn/Fe$]$ & $[$Ni/Fe$]$ & $[$Zn/Fe$]$ \\
         &   II  &   I   &   I   &   II  &   I   &   I   &   I   \\
\hline
  224930 &       &  0.18 &  0.04 &       &       &       &       \\
    3567 & -0.05 & -0.02 & -0.11 &  0.05 & -0.48 &       & -0.06 \\
    3628 &       & -0.05 &  0.02 &       &       &       &       \\
-35~0360 &  0.09 &  0.20 &  0.00 & -0.05 & -0.37 & -0.03 &       \\
    6582 &       &  0.15 &  0.00 &       &       & -0.06 &       \\
    9430 &       &  0.05 &  0.04 &       &       & -0.09 &       \\
-61~0282 & -0.15 & -0.09 & -0.09 &  0.03 & -0.47 & -0.04 & -0.03 \\
   10607 &  0.34 &  0.04 & -0.04 & -0.04 & -0.42 & -0.06 &  0.36 \\
+29~0366 &  0.11 &  0.02 & -0.08 & -0.06 & -0.31 & -0.02 & -0.04 \\
-01~0306 &       &  0.02 & -0.00 &       &       & -0.11 &       \\
   15096 &       &  0.01 & -0.16 &       &       & -0.04 &       \\
   16397 &       & -0.04 & -0.08 &       &       & -0.07 &       \\
   17288 &       &       & -0.00 &  0.03 &       &  0.03 &       \\
   17820 &       &  0.03 & -0.04 &  0.03 &       &  0.03 &       \\
   18907 &       &  0.03 & -0.07 &       &       & -0.00 &       \\
   19445 &  0.01 &       & -0.10 &  0.05 & -0.56 & -0.02 &  0.09 \\
   20512 &       &  0.01 & -0.03 &       &       &  0.00 &       \\
-47~1087 &       &       &  0.00 &  0.00 &       &  0.05 &       \\
   22879 &       & -0.05 & -0.05 &  0.01 &       & -0.01 &       \\
   23439 &  0.06 &  0.16 & -0.01 &  0.15 & -0.36 & -0.04 &  0.19 \\
   24616 &       &  0.15 & -0.12 &       &       & -0.02 &       \\
   25704 & -0.01 &  0.19 & -0.13 & -0.06 & -0.37 & -0.01 & -0.01 \\
   25329 &       &  0.33 &  0.07 &       &       &  0.03 &       \\
   25673 &       & -0.16 &       &       &       &  0.00 &       \\
  284248 &       &       & -0.21 &       &       & -0.10 &       \\
   29907 & -0.16 & -0.01 &  0.10 &  0.27 & -0.47 & -0.14 &  0.19 \\
  280067 &       &  0.17 & -0.02 &       &       & -0.02 &       \\
   29400 &  0.18 &  0.25 &  0.03 &  0.13 & -0.10 & -0.02 &  0.08 \\
   31128 & -0.02 &  0.54 &  0.07 &  0.13 & -0.19 &  0.06 &  0.15 \\
  241253 &  0.04 &       & -0.13 & -0.06 & -0.38 & -0.04 & -0.01 \\
   34328 & -0.09 &  0.05 & -0.11 &  0.18 & -0.46 & -0.08 &  0.24 \\
   36283 &  0.21 &  0.14 & -0.03 &  0.04 & -0.15 & -0.02 &  0.17 \\
+12~0853 &       &       & -0.01 &  0.14 &       &  0.04 &       \\
   40057 &       &  0.04 & -0.08 &       &       & -0.14 &       \\
   45205 &       & -0.03 & -0.08 &       &       & -0.02 &       \\
   46341 &       & -0.17 &  0.08 &       &       & -0.13 &       \\
-25~3416 &       &  0.29 & -0.02 &       &       & -0.06 &       \\
-33~3337 &       &       & -0.14 &       &       & -0.01 &       \\
   51754 &       &       &  0.11 &       &       &  0.01 &       \\
   53545 &       & -0.04 &       &       &       & -0.03 &       \\
-57~1633 &       &       & -0.10 & -0.10 &       & -0.18 &       \\
   53871 &       &  0.04 & -0.03 &       &       & -0.10 &       \\
+17~1524 &  0.09 &  0.30 &  0.04 &  0.05 & -0.01 &  0.02 &  0.11 \\
   59374 &       &  0.10 & -0.07 &       &       & -0.06 &       \\
-45~3283 &       &       & -0.07 &  0.11 &       & -0.12 &       \\
   60319 &       &  0.11 & -0.04 & -0.00 &       &  0.03 &       \\
   64090 & -0.07 & -0.04 & -0.08 & -0.09 & -0.42 & -0.18 &  0.07 \\
   64606 &       &  0.19 & -0.02 &       &       & -0.00 &       \\
+23~3511 &       &  0.01 & -0.16 &       &       & -0.10 &       \\
   74000 &       &       & -0.08 &       &       &       &       \\
\hline
\end{tabular}
\normalsize
\label{t:abund1}
\end{table*}

\addtocounter{table}{-1}

\begin{table*}
\caption{Abundances}
\scriptsize
\begin{tabular}{rrrrrrrr}
\hline
HD/DM & $[$Sc/Fe$]$ & $[$V/Fe$]$ & $[$Cr/Fe$]$ & $[$Cr/Fe$]$ & $[$Mn/Fe$]$ & $[$Ni/Fe$]$ & $[$Zn/Fe$]$ \\
         &   II  &   I   &   I   &   II  &   I   &   I   &   I   \\
\hline
   75530 &  0.21 &  0.30 &  0.05 &  0.06 & -0.11 & -0.03 &  0.07 \\
   76932 &       & -0.06 & -0.06 & -0.08 &       & -0.00 &       \\
   76910 &       &  0.12 & -0.11 &       &       & -0.05 &       \\
-03~2525 &       &       & -0.05 &       &       &  0.26 &       \\
   78737 &       &       & -0.11 &       &       &       &       \\
-80~0328 &  0.07 &       & -0.06 & -0.22 &       & -0.06 &  0.07 \\
   83220 &       &       & -0.15 & -0.09 &       & -0.04 &       \\
   83888 &       &  0.00 & -0.04 &       &       & -0.14 &       \\
+09~2242 &       &  0.01 & -0.03 &       &       & -0.15 &       \\
   84937 &  0.04 &       & -0.09 & -0.11 & -0.50 &  0.01 &  0.30 \\
   88725 &       & -0.02 &  0.02 &       &       & -0.07 &       \\
   91345 & -0.03 & -0.07 &  0.08 &  0.17 & -0.35 & -0.02 &  0.27 \\
+29~2091 &       &       & -0.15 &       &       &       &       \\
   94028 & -0.01 &       & -0.12 &  0.07 & -0.38 &  0.01 &  0.17 \\
   97320 &  0.11 &  0.04 & -0.11 &  0.10 & -0.31 & -0.02 &  0.27 \\
   97916 &       &       & -0.15 &       &       & -0.03 &       \\
  103095 &  0.00 &  0.12 & -0.06 &  0.07 & -0.35 & -0.08 &  0.16 \\
  105755 &       &       & -0.02 &       &       & -0.06 &       \\
  106038 &       &       & -0.06 &  0.07 &       &  0.18 &       \\
  106516 &       &  0.09 & -0.10 & -0.06 &       &  0.00 &       \\
  108076 &       & -0.06 & -0.02 &       &       &  0.01 &       \\
  108177 & -0.07 & -0.18 & -0.16 & -0.09 &       & -0.03 &       \\
  111980 &  0.12 & -0.02 & -0.12 & -0.05 & -0.47 & -0.03 &  0.13 \\
 113083A &       &       & -0.13 & -0.07 &       & -0.10 &       \\
 113083B &       &       & -0.09 & -0.05 &       & -0.13 &       \\
  113679 &       &       & -0.00 & -0.00 &       &  0.05 &       \\
+33~2300 &       & -0.07 & -0.00 &       &       & -0.16 &       \\
  114762 &       &  0.09 & -0.08 &       &       & -0.04 &       \\
  116064 & -0.06 &       & -0.11 & -0.14 &       &       &       \\
  116316 &       &  0.08 & -0.07 &       &       & -0.04 &       \\
  118659 &       &  0.08 & -0.02 &       &       & -0.10 &       \\
  119173 &       & -0.09 & -0.09 &       &       & -0.12 &       \\
  120559 &  0.16 &  0.09 & -0.01 &  0.13 & -0.21 &  0.07 &  0.38 \\
  121004 &  0.11 & -0.05 & -0.03 &  0.07 & -0.23 &  0.03 &  0.35 \\
  123710 &       & -0.11 & -0.03 &       &       & -0.08 &       \\
  126681 &  0.16 & -0.08 & -0.06 & -0.09 & -0.48 & -0.05 &       \\
  129515 &       & -0.02 & -0.04 &       &       & -0.12 &       \\
  129392 &       &  0.13 & -0.05 &       &       & -0.12 &       \\
  129518 &       &  0.00 & -0.06 &       &       & -0.13 &       \\
+26~2606 &       &       & -0.09 &       &       &       &       \\
  132475 &  0.19 & -0.04 & -0.14 &  0.05 & -0.30 &  0.02 &       \\
  134113 &       &  0.05 & -0.08 &       &       & -0.09 &       \\
  134088 &       &  0.02 & -0.06 &       &       & -0.04 &       \\
  134169 &  0.03 &  0.05 & -0.06 &  0.09 & -0.13 & -0.01 &  0.06 \\
  134439 & -0.07 & -0.12 & -0.06 &  0.17 & -0.41 & -0.22 & -0.13 \\
  134440 & -0.46 & -0.02 & -0.16 &  0.20 & -0.44 & -0.11 &       \\
  140283 &       &       & -0.14 &  0.17 & -0.61 &  0.11 &       \\
  142575 &       &  0.17 & -0.11 &       &       &  0.03 &       \\
+42~2667 &       &  0.00 & -0.12 &       &       & -0.07 &       \\
  145417 & -0.02 &  0.15 &  0.06 & -0.02 & -0.34 & -0.03 &       \\
\hline
\end{tabular}
\normalsize
\end{table*}

\addtocounter{table}{-1}

\begin{table*}
\caption{Abundances}
\scriptsize
\begin{tabular}{rrrrrrrr}
\hline
HD/DM & $[$Sc/Fe$]$ & $[$V/Fe$]$ & $[$Cr/Fe$]$ & $[$Cr/Fe$]$ & $[$Mn/Fe$]$ & $[$Ni/Fe$]$ & $[$Zn/Fe$]$ \\
         &   II  &   I   &   I   &   II  &   I   &   I   &   I   \\
\hline
  148816 &       &  0.10 & -0.01 &       &       & -0.04 &       \\
  149414 &       &  0.13 &  0.01 &       &       & -0.02 &       \\
  149996 &       &  0.03 &  0.02 &       &       & -0.01 &       \\
  157466 &       & -0.05 & -0.08 &       &       & -0.11 &       \\
+31~3025 &  0.26 &  0.32 &  0.04 &  0.03 & -0.12 & -0.02 &  0.08 \\
  158226 &       &  0.10 & -0.00 &       &       &  0.01 &       \\
  158809 &       &  0.04 &  0.02 &       &       & -0.07 &       \\
  159482 &  0.01 &  0.09 & -0.04 & -0.02 & -0.35 &  0.03 &  0.23 \\
  160693 &       & -0.07 & -0.04 &       &       & -0.11 &       \\
+02~3375 & -0.05 &       &  0.05 & -0.06 & -0.41 &  0.45 &  0.36 \\
  163810 &       &  0.04 & -0.00 &       &       & -0.11 &       \\
  163799 &       &  0.09 & -0.05 &       &       & -0.04 &       \\
+05~3640 &       &  0.15 & -0.04 &  0.16 & -0.38 &  0.05 &  0.19 \\
  166913 &  0.08 &       & -0.12 & -0.05 & -0.35 & -0.01 &  0.13 \\
  171620 &       &  0.06 & -0.05 &       &       & -0.05 &       \\
  174912 &       & -0.07 & -0.06 &       &       & -0.08 &       \\
  175179 &       & -0.02 & -0.06 &       &       & -0.07 &       \\
  179626 &       &  0.16 & -0.06 &       &       & -0.06 &       \\
  181743 & -0.14 &       & -0.08 &  0.20 & -0.42 &       &  0.01 \\
  184499 &       &  0.04 & -0.04 &       &       & -0.06 &       \\
  186379 &       & -0.07 & -0.03 &       &       & -0.07 &       \\
  188510 & -0.18 &  0.12 & -0.01 &  0.07 & -0.40 & -0.18 &  0.04 \\
  189558 & -0.01 &  0.13 & -0.10 & -0.09 & -0.40 & -0.03 &  0.34 \\
+42~3607 &       &       & -0.10 &       &       &  0.20 &  0.10 \\
+23~3912 &  0.06 &  0.02 & -0.07 &  0.06 & -0.42 & -0.06 &  0.05 \\
  192718 &       &  0.07 & -0.03 &       &       & -0.01 &       \\
  193901 & -0.12 & -0.10 & -0.10 & -0.07 & -0.41 & -0.20 &       \\
  194598 & -0.04 &  0.03 & -0.07 &  0.03 & -0.29 & -0.05 & -0.10 \\
  195633 &       &  0.01 & -0.06 &       &       & -0.05 &       \\
  195987 &       &  0.20 & -0.07 &       &       &  0.13 &       \\
  196892 &  0.13 &  0.04 & -0.11 & -0.04 & -0.35 &  0.01 &  0.16 \\
+41~3931 &       &  0.12 & -0.11 &       &       & -0.10 &       \\
+33~4117 &  0.06 &  0.11 &  0.03 &  0.01 & -0.02 & -0.07 & -0.01 \\
+19~4601 &  0.16 &  0.12 &  0.00 &  0.04 & -0.22 & -0.04 &  0.08 \\
  201891 &  0.17 & -0.04 & -0.13 & -0.07 & -0.28 & -0.04 &  0.01 \\
  201889 &       &  0.22 & -0.01 &       &       & -0.07 &       \\
  204155 &  0.05 &  0.05 & -0.02 & -0.06 & -0.30 & -0.01 &  0.09 \\
  205650 & -0.02 &  0.03 & -0.15 & -0.15 & -0.44 & -0.04 &       \\
+59~2407 &       &  0.24 & -0.04 &       & -0.34 &  0.13 & -0.05 \\
+22~4454 &       &  0.15 &  0.09 &       &       & -0.04 &       \\
  207978 &       &  0.14 & -0.11 &       &       & -0.03 &       \\
+11~4725 &       &  0.12 &  0.01 &       &       & -0.08 &       \\
+17~4708 & -0.11 &  0.17 & -0.11 & -0.08 & -0.24 & -0.21 &  0.14 \\
  211998 &  0.00 & -0.08 &  0.12 & -0.05 & -0.46 & -0.08 &  0.26 \\
  218502 &       &       & -0.08 &       &       &  0.01 &       \\
 219175A & -0.07 & -0.24 & -0.07 & -0.07 & -0.22 & -0.02 &  0.15 \\
 219175B & -0.04 & -0.34 & -0.02 &  0.30 & -0.19 &  0.07 &  0.39 \\
+33~4707 &       & -0.12 & -0.28 &       &       &  0.09 &       \\
  221377 &       &  0.11 & -0.10 &       &       & -0.03 &       \\
  222794 &       & -0.01 & -0.02 &       &       &  0.01 &       \\
\hline
\end{tabular}
\normalsize
\end{table*}

\begin{table*}
\caption{Solar analysis}
\begin{tabular}{lrlrrrr}
\hline  
Species & N. lines & {\it gf} source &\multicolumn{2}{c}{Solar Abundance} & 
Meteorites & Notes \\ 
      &     &                             & HM   & Kurucz &    &       \\
\hline
[O~I] &   1 & Allende-Prieto et al. 2002  & 8.81 & 8.76 &      &       \\ %
O~I   &   3 & Bi\'emont et al. 1991a      & 8.83 & 8.79 &      &       \\ %
Na~I  &   7 & NIST                        & 6.33 & 6.21 & 6.31 &       \\ %
Mg~I  &   4 & See Appendix                & 7.52 & 7.43 & 7.58 &       \\ %
Si~I  &  14 & Garz 1973                   & 7.61 & 7.53 & 7.55 &       \\ %
Ca~I  &  24 & Smith \& Raggett 1981       & 6.39 & 6.27 & 6.34 &       \\ %
Sc~II &  12 & NIST                        & 3.16 & 3.13 & 3.09 & HFS   \\ %
Ti~I  &  41 & Oxford group                & 5.10 & 5.00 & 4.93 &       \\ %
Ti~II &  20 & Bizzarri et al. 1993        & 5.11 & 5.07 & 4.93 &       \\ %
V~I   &  18 & Whaling et al. 1985         & 4.07 & 3.97 & 4.02 & HFS   \\ %
Cr~I  &  60 & Oxford group                & 5.76 & 5.67 & 5.68 &       \\ %
Cr~II &  21 & Solar                       & 5.75 & 5.71 & 5.68 &       \\ %
Mn~I  &  11 & Booth et al. 1984           & 5.42 & 5.34 & 5.53 & HFS   \\ %
Fe~I  & 173 & Oxford group                & 7.62 & 7.54 & 7.51 &       \\ %
      &     & Bard et al. 1991            &      &      &      &       \\ %
      &     & Bard \& Kock 1994           &      &      &      &       \\ %
      &     & O'Brian et al. 1991         &      &      &      &       \\ %
Fe~II &  41 & Holweger et al. 1990        & 7.50 & 7.49 & 7.51 &       \\ %
      &     & Bi\'emont et al. 1991b      &      &      &      &       \\ %
      &     & Hannaford et al. 1992       &      &      &      &       \\ %
      &     & Blackwell et al. 1980       &      &      &      & +0.18 \\ %
Ni~I  &  47 & Solar                       & 6.36 & 6.28 & 6.25 &       \\ %
Zn~I  &   4 & Bi\'emont \& Godefroid 1980 & 4.63 & 4.59 & 4.65 &       \\ %
\hline  
\end{tabular}
\label{t:solar}
\end{table*}

\section{Abundances}

The abundances for a number of elements determined from the original spectra
are listed in Tables~\ref{t:abund0} and \ref{t:abund1}. We do not give here
abundances for neutron capture elements, that will be considered in a paper
in preparation (Fran\c cois et al. 2003). When equivalent widths from
different sources were available for a star, the abundances listed in these
tables are the average of those derived using equivalent widths from different
sources (each one analyzed as an independent spectrum), averaged with a weight
equal to the number of lines used in each analysis. Finally, hereinafter
$\alpha-$element is the average of Mg, Si, Ca, and Ti.

Abundances given throughout this paper were computed with respect to a solar
abundance analysis done using the same procedure adopted throughout this
paper, and the solar model atmosphere computed by Kurucz (1994) (see
Table~\ref{t:solar}); 
for reference, also the abundances obtained using
the Holweger \& M\"uller (1974) model atmosphere are given. 
We think that this differential procedure minimizes
errors in the analysis of stars similar to the Sun. For comparison, meteoritic
abundances (from Anders \& Grevesse 1989) are also listed in
Table~\ref{t:solar}.

Sources of oscillator strengths are also given in Table~\ref{t:solar}.
We tried to use laboratory and theoretical oscillator strengths for those
lines for which accurate results (errors below 0.05 dex) exist. For the
remaining lines, they were derived from an inverse solar analysis using
the Holweger \& M\"uller (1974) empirical model atmosphere, and elemental
abundances given by lines with theoretical/laboratory oscillator strengths.

Whenever possible (permitted transitions with not too large effective quantum
number between S, P, D and F levels), collisional damping was considered using
coefficients from Barklem et al. (2000), that are the best theoretical models
available at present for most transitions. For a few lines, data missing in
this very extensive tabulation were computed using the WIDTHCOMP program
written by Barklem, and available on the web (see Barklem et al. 1998). For
several Mg I lines, which have large effective quantum number, empirical
enhancement factors to classical damping were obtained by fitting the line
profiles in the Kurucz et al. (1984) Solar Spectrum. For the remaining
transitions, we considered classical damping computed with the Uns\"old (1955)
formula, multiplied by an enhancement factor $E$\ to the $C_6$\ constant given
by:
\begin{equation}
\log E = (0.381\pm 0.017) EP - (0.88\pm 0.33),
\end{equation}
where $EP$\ is the line excitation potential (in eV). This formula was 
obtained from several hundreds Fe~I line for which accurate collisional
damping parameters were available (the formula was obtained at T=5000 K;
the temperature dependence of collisional damping constant given by the
Uns\"old formula being slightly different from that obtained using more
accurate approaches).

O and Na abundances include corrections for departures from LTE computed
according the precepts by Gratton et al. (1999). Most O abundances are
obtained from the IR triplet. The forbidden lines have been measured in a few
field stars; they are listed separately.

Corrections for hyperfine structure, due to non-zero nuclear magnetic moment,
have been applied to abundances for Sc, V, and Mn.
Data were from NIST database, Booth et al. (1983), Whaling et al. (1985),
and Prochaska \& McMillan (2000).

Abundances for dominant species (O, Ti II, Sc II, and Cr II) were compared
with abundances from Fe II, to reduce the impact of uncertainties in the
surface gravities.

\subsection{Error analysis}

\begin{table*}
\caption{Sensitivity of abundances on errors in the atmospheric parameters}
\begin{tabular}{lrrrrrrr}
\hline  
Element & N. lines & $\Delta$T & $\Delta \log{g}$ & $\Delta [$A/H$]$ & $\Delta v_t$ & $\Delta EW$ &
total\\
        & median & +100~K    & +0.3~dex         & +0.2~dex         & +0.2~km/s    & m\AA & dex \\
\hline  
$[$Fe/H$]$I   & 46 &  ~0.079 & $-$0.015 &   ~0.011 & $-$0.019 & 0.013 & 0.051 \\
$[$Fe/H$]$II  & 17 &$-$0.077 &   ~0.133 &   ~0.001 &    0.009 & 0.022 & 0.066 \\
\\
$[$O/Fe$]$I(f)&  1 &   0.035 & $-$0.003 &    0.024 &   ~0.008 & 0.090 & 0.096 \\
$[$O/Fe$]$I(p)&  3 &$-$0.077 & $-$0.003 & $-$0.025 &   ~0.007 & 0.052 & 0.069 \\
$[$Na/Fe$]$I  &  4 &$-$0.038 & $-$0.007 & $-$0.008 &   ~0.013 & 0.045 & 0.054 \\
$[$Mg/Fe$]$I  &  4 &$-$0.032 & $-$0.004 & $-$0.005 &   ~0.016 & 0.045 & 0.055 \\
$[$Si/Fe$]$I  & 11 &$-$0.054 &   ~0.049 & $-$0.008 &   ~0.014 & 0.027 & 0.048 \\
$[$Ca/Fe$]$I  & 12 &$-$0.025 &   ~0.002 & $-$0.004 &   ~0.006 & 0.026 & 0.033 \\
$[$Sc/Fe$]$II &  3 &  ~0.029 & $-$0.003 &   ~0.006 &   ~0.003 & 0.052 & 0.058 \\
$[$Ti/Fe$]$I  & 13 &$-$0.003 &   ~0.022 & $-$0.001 &   ~0.005 & 0.025 & 0.030 \\
$[$Ti/Fe$]$II &  9 &  ~0.037 & $-$0.031 &   ~0.007 & $-$0.014 & 0.030 & 0.048 \\
$[\alpha/$Fe$]$&49 &$-$0.023 &   ~0.011 & $-$0.003 &   ~0.008 & 0.013 & 0.025 \\
$[$Cr/Fe$]$I  & 10 &$-$0.014 &   ~0.030 & $-$0.001 &   ~0.012 & 0.028 & 0.038 \\
$[$Cr/Fe$]$II &  4 &$-$0.008 & $-$0.002 & $-$0.003 &   ~0.004 & 0.045 & 0.073 \\
$[$Mn/Fe$]$I  &  6 &$-$0.018 &   ~0.031 &   ~0.001 &   ~0.009 & 0.037 & 0.043 \\
$[$Ni/Fe$]$I  & 13 &$-$0.016 &   ~0.006 &   ~0.000 &   ~0.013 & 0.025 & 0.035 \\
$[$Zn/Fe$]$I  &  2 &$-$0.050 &   ~0.081 & $-$0.002 &   ~0.008 & 0.064 & 0.076 \\
\hline  
\end{tabular}
\label{t:err}
\end{table*}

Relevant data for the error analysis are given in Table~\ref{t:err}. Col.~2
gives the median number of equivalent widths for this element used in the
stellar analysis for the whole sample; Cols~3 to 6 give the sensitivities
of our abundances on changes in the adopted atmospheric parameters. They were
obtained by changing one at a time individual parameters for a typical program
star. Col.~7 lists the expected contribution to errors due to the number of
lines measured in each star (given by the typical error for an individual line
divided by the square root of the median number of lines used): these are
typical values, since the actual number of lines used in the analysis varies
from star-to-star. Finally, the last column gives the typical error bar for
abundances in each individual star: it is obtained by summing quadratically
errors due to atmospheric parameters and to the equivalent widths. When
estimating this value, we assumed typical errors of $\pm 50$~K in $T_{\rm
eff}$, $\pm 0.1$~dex in $\log g$, $\pm 0.1$~dex in [Fe/H], $\pm 0.3$~km/s in
$v_t$, and 0.09~dex in abundances derived from an individual line (this is the
r.m.s. average of the scatter for individual Fe I lines, obtained over all the
stars).

\begin{figure*}
\centering
\includegraphics[width=13cm]{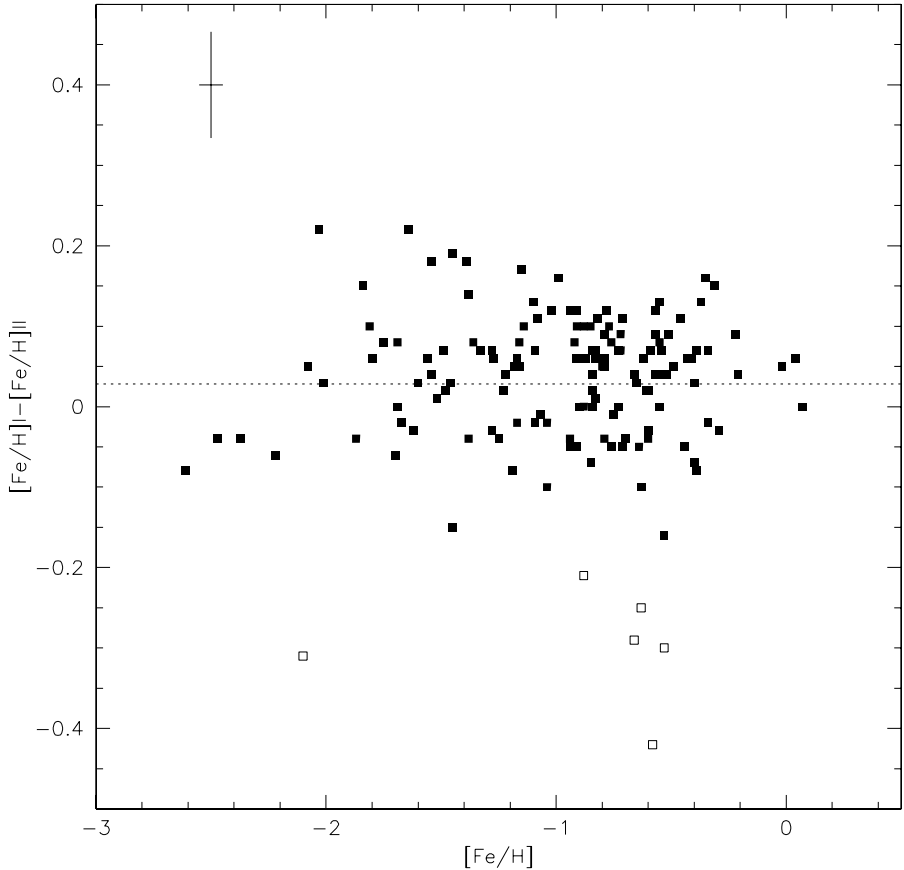}
\caption{ Run of the differences between abundances obtained from Fe~I and
Fe~II lines as a function of overall metal abundance [Fe/H]. The stars
considered to be outliers using a 3~$\sigma$\ clipping procedure are marked
with open symbols; the other stars are plotted as filled symbols. The dashed
line is at at the average value of 0.028 dex (see text). Typical
error bars for an individual star is also shown }
\label{f:fe2}
\end{figure*}  

In the remaining part of this paper we will briefly comment about the
abundances of Fe and O. A full discussion of the present abundances will be
given in the second paper of this series.

\subsection{Fe abundances}

Since gravities are not derived from the equilibrium of ionization, the
difference between abundances provided by neutral and singly ionized Fe lines
can be used to test the equilibrium of ionization. Once the offset of 0.05 dex
found in solar Fe abundances using the same line parameters used throughout
this paper is subtracted, on average, this difference is $0.028\pm 0.008$~dex,
in the sense that abundances given by singly ionized Fe lines are smaller
(r.m.s.=0.098 dex, 146 stars, since we have not any abundance from singly
ionized Fe lines in HD51754, HD74000, BD+29~2091, BD+42~3607). If a few
outliers are eliminated using a clipping procedure at 3~$\sigma$, this
difference is $0.041\pm 0.006$~dex (r.m.s.=0.071~dex, 140 stars); the outliers
are HD15096, HD280067, BD$-$3~2525, HD195987, HD219175B, and BD+33~4707
(note that these outliers do not coincide with those identified when
comparing our temperatures with those given by Alonso et al. 1996). In all
these cases abundances from Fe II lines are significantly larger than those
from Fe I lines, suggesting that temperatures for these stars are higher than
assumed in our analysis. Note that $b-y$\ colors are missing for HD280067; it
is a metal-rich star, and the temperature derived from $B-V$\ color alone
might have a rather large error. Four other outliers are known binaries
(HD15096, BD-3~2525, HD195987, and HD219175B), and the secondary might affect
both the magnitude and the color of the star in a significant way. Finally,
it is possible that some of these stars are reddened. Given the uncertainties
present in their analyses, these six stars will not be considered in the
following discussion.

If further known or suspected binaries or reddened stars are excluded from the
comparison, the average difference between abundances from Fe~I and Fe~II
lines (in the same sense as above) is $0.055\pm 0.007$~dex (r.m.s.=0.062, 83
stars). We did not found any significant trend of these residuals with
temperature, gravity, or overall metal abundance. The observed r.m.s scatter
agrees fairly well with the expected value of 0.066~dex (see last column of
Table~\ref{t:err}).

While statistically significant, the average offset between abundances from
neutral and singly ionized Fe lines is clearly small. It might be a result of
small errors ($\sim 50$~K) in the effective temperature scale adopted
throughout this paper (however, our scale agrees with 
the temperature scale by Alonso et al., that is the
best currently available for metal-poor stars), or of systematic deviations
between the adopted model atmospheres and the real ones (slightly different
from those observed for the Sun). Nissen et al. (2002) have very recently
studied the effect of granulation on the formation of Fe II lines; they showed
that in their 3-D models, Fe II lines with excitation of $\sim 3$~eV (a
typical value for lines in our line list) are expected to be weaker than in
1-D models. The effect is expected to be roughly proportional to overall
metallicity (it should be almost negligible in the Sun), and the expected 
corrections (a few hundredths of a dex) match well the observed average
difference found here between abundances given by Fe I and Fe II lines.

In any case, we did not find any clear evidence for significant departures
from LTE in the formation of Fe lines: as a matter of fact, the difference we
finally have, if any, is in the opposite direction with respect to prediction
from Fe overionization. We cannot then confirm the significant Fe
overionization (at $\sim 0.2$~dex in metal-poor dwarfs) claimed by Idiart \&
Thevenin (2000).

\subsection{Comparison with literature [Fe/H] values}

\begin{figure}
\centering
\includegraphics[width=8.8cm]{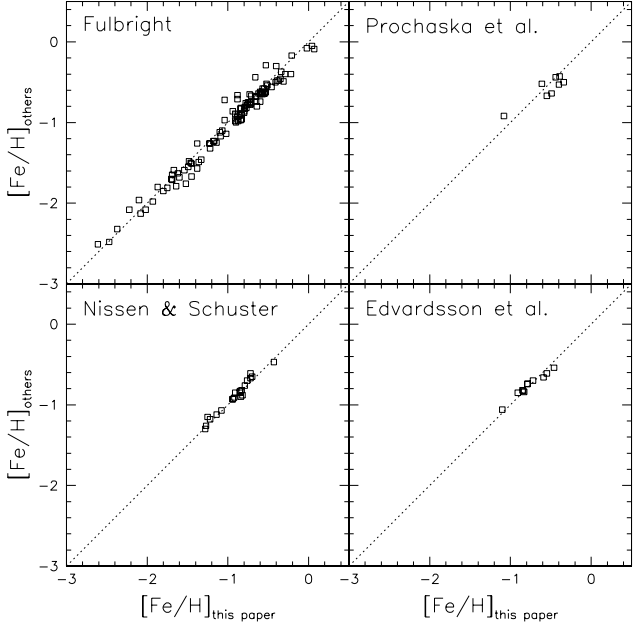}
\caption{ Comparison between [Fe/H] values determined in this paper and
those from the literature: Fulbright (2000), Nissen \& Schuster (1997),
Prochaska et al. (2000), and Edvardsson et al. (1993) }
\label{f:compabu}
\end{figure}  

\begin{table}
\caption{Comparison of [Fe/H] values}
\begin{tabular}{lccccl}
\hline  
Data set   & n. stars & $\Delta[$Fe/H$]$ & rms & Remark       \\
\hline  
Fulbright  & 110 & $-0.034\pm 0.009$ & 0.093 & all            \\
           & 100 & $-0.049\pm 0.006$ & 0.062 & $-$10 outliers \\
Nissen     &  21 & $+0.020\pm 0.010$ & 0.044 & all            \\
Prochaska  &   8 & $-0.044\pm 0.042$ & 0.119 & all            \\
Edvardsson &  11 & $+0.004\pm 0.015$ & 0.051 & all            \\
\hline  
\end{tabular}
\label{t:compabu}
\end{table}

In Fig.~\ref{f:compabu} we compare the [Fe/H] values determined in this paper
with those from the literature. Table~\ref{t:compabu} lists average differences
(in the sense other studies minus us) computed using stars in common.
The agreement is generally good. However,
a close insight reveals some small systematic differences. On average, our
abundances are slightly larger than those of Fulbright (2000) and Prochaska
et al. (2000); in both cases the scatter is not negligible, although in the
case of Fulbright paper, it is reduces in a significant way by eliminating ten
discrepant cases using a 2.5~sigma clipping procedure (8 out of 10 of these
discrepant cases are binaries). On the other side, our abundances are on
average slightly lower than those of Nissen \& Schuster (1997); the difference
is however only 0.02 dex, and the star-to-star agreement is in this case
excellent. Finally there is a very good agreement with the abundances by
Edvardsson et al. (1993). It should be noted that this is the only really
independent data set here considered, since we used the $EW$s listed in the
other studies to compute our abundances.

\subsection{Oxygen}

We have derived Oxygen abundances using both the IR triplet lines at
7771-74~\AA\ (for 68 stars), and the forbidden line at 6300.304~\AA\ (for 22
stars). In the second case, for 11 stars we used the $EW$s measured by Nissen et
al. (2002), including the correction for blending with the Ni line at
6300.399~\AA; for two further stars (HD3567 and HD97320), we were able to
measure again the $EW$s using our own spectra: these $EW$s turned out (perhaps
for some lucky coincidence, since our data are of lower quality with respect to 
those of Nissen et al.) to be exactly identical to those given by Nissen et al.
However, we have to keep in mind that the forbidden line is always very weak in
our stars, and that abundances are very sensitive on the correct location of the
continuum level and to the noise.

Non-LTE corrections are required to obtain abundances from the permitted lines
(the forbidden lines form in LTE). These corrections were considered,
following the approach by Gratton et al. (1999); however, since there is
growing evidence that collisional cross sections with HI atoms computed using
the Drawin approach are overestimated in these computations, we corrected them
in order to be consistent with those of Nissen et al. (2002). Anyhow, the
non-LTE corrections for the permitted lines are not very large in the program
stars ($<0.25$~dex), in agreement with results by several other authors (see
e.g. Kiselman, 2001). It is useful to note that when updated $gf$'s are used
and non-LTE corrections are included, permitted and forbidden lines give very
similar O abundances when the Kurucz model atmosphere are used in the solar
analysis (see Table~\ref{t:solar}).

We may compare abundances provided by (high excitation) permitted lines and
(low excitation) forbidden ones in 20 stars. On average, the difference (in
the sense permitted-forbidden) is $0.12\pm 0.04$~dex, with an r.m.s. of 0.18
dex for individual stars. Both the mean difference and the scatter of
individual abundances are quite large (the expected value for the r.m.s is 0.12
dex). However, most of this uncertainty comes from those stars which are 
either known binaries or are likely to be significantly reddened (we remind
here that we did not correct colors for reddening when deriving parameters
for the stars). In both cases, temperatures are likely underestimated for these
stars, resulting in larger abundances from the permitted lines, and lower from
the forbidden ones. When these stars are excluded from the comparison, the
mean difference (based now on 10 stars) is $0.02\pm 0.04$~dex, with an r.m.s.
of 0.13~dex, in good agreement with the expected error bar. This agreement is
comfortable, and supports the result obtained by Nissen et al. (2002) on a
somewhat smaller sample.

These abundances have been obtained using the Kurucz 1-D model atmospheres; as
we have seen when commenting the Fe abundances, these models might not be an
adequate description of the real model atmospheres when very accurate
abundances are considered. The impact of granulation on the formation of
Oxygen lines has been considered in detail by Nissen et al. (2002). They
suggest that application of 3-D model atmospheres reduces the abundances from
the forbidden line more than those from the permitted lines. Unless the non-LTE
corrections to the OI lines are enhanced in 3-D model atmospheres, this would
create a difference between abundances from permitted and forbidden lines in
metal-poor dwarfs. On the other hand, the structure of model atmosphere may be
simply incorrect: to show the impact of such a difference, we may consider the
abundances obtained replacing the Kurucz model atmospheres without
overshooting (used throughout this paper), with those computed with the
overshooting option switched on. Typically, differences between abundances
from permitted and forbidden Oxygen lines would be reduced by 0.04 dex
(however the difference between abundances from Fe I and Fe II lines would be
further increased by 0.02 dex).

\begin{acknowledgements}
This research has made use of the SIMBAD data base, operated at CDS,
Strasbourg, France; of the NIST database, operated by the National
Institute of Standards and Technology; and of VALD database. 
We wish to thank L. Pasquini, V. Hill, M. Centurion,
P. Bonifacio, and C. Sneden for help during the observations,
F. Primas for having provided two reduced spectra,
and P. Bertelli for useful comments. This research was funded by
COFIN 2001028897 by Ministero Universit\`a e Ricerca Scientifica,
Italy
\end{acknowledgements}

\begin{table*}
\caption{Lifetimes for singlet Mg I levels}
\begin{tabular}{lccccc}
\hline  
Level     & Chantepie &   J\"onsson  & Chang  &   FF   &    Adopted     \\
          &   (ns)    &     (ns)     &  (ns)  &  (ns)  &      (ns)      \\
\hline  
$4s^1S$   &  $44\pm 5$ &  $47\pm  3$ &  45.8  &  45    &  $46.2\pm 2.6$ \\
$5s^1S$   & $102\pm 5$ & $100\pm  5$ & 100~~  & 102    & $101.0\pm 3.5$ \\
$6s^1S$   & $215\pm 7$ & $211\pm 12$ & 196~~  & 204    & $214.0\pm 6.0$ \\
$4p^1P^0$ &            &             &  14.3  &  14    &      14.3      \\
$6p^1P^0$ &            &             & 121    & 116    &     121        \\
$3d^1D$   &  $72\pm 4$ &  $81\pm  6$ &  79.5  &  45    &  $74.8\pm 3.3$ \\
$4d^1D$   &  $53\pm 3$ &  $57\pm  3$ &  52.4  &  39    &  $55.0\pm 2.1$ \\
$5d^1D$   &  $45\pm 1$ &  $50\pm  4$ &  42.6  &  37    &  $45.3\pm 1.0$ \\
$6d^1D$   &  $52\pm 1$ &  $54\pm  3$ &  51.8  &  50    &  $52.2\pm 0.9$ \\
$7d^1D$   &  $71\pm 2$ &  $70\pm  6$ &  69.2  &  65    &  $70.9\pm 1.9$ \\
$8d^1D$   &  $94\pm 4$ &  $93\pm  7$ &  94.1  &  92    &  $93.8\pm 3.5$ \\
\hline  
\end{tabular}
\label{t:mglifetimes1}
\end{table*}

\begin{table*}
\caption{Lifetimes for triplet Mg I levels}
\begin{tabular}{lcccccc}
\hline  
Level     & Schaefer &  Kwiatkowski  & Andersen & Chang  &   FF   &   Adopted     \\
          &   (ns)    &     (ns)     &   (ns)   &  (ns)  &  (ns)  &     (ns)      \\
\hline  
$4s^3S$   &$14.8\pm 0.7$& $9.7\pm 0.6$&$10.1\pm 0.8$&   9.98 &   9.86 &  $9.8\pm 0.3$ \\
$5s^3S$   &$25.6\pm 2.1$&             &             &  27.5  &  26.8  & $25.6\pm 2.1$ \\
$6s^3S$   &$52.1\pm 6.0$&$51.8\pm 3.0$&             &  58.9  &  57.2  & $51.8\pm 3.0$ \\
$5p^3P^0$ &             &             &             & 256    & 269    &    256        \\
$6p^3P^0$ &             &             &             & 590    & 620    &    590        \\
$5d^3D$   &             &$34.1\pm 1.5$&             &  33.3  &  34.5  & $34.1\pm 1.5$ \\
$6d^3D$   &             &$55.7\pm 3.0$&             &  59.2  &  66.9  & $55.7\pm 3.0$ \\
$7d^3D$   &             &$91.5\pm 5.0$&             &  96.0  & 112    & $91.5\pm 5.0$ \\
$8d^3D$   &             &             &             &        & 173    &    173        \\
\hline  
\end{tabular}
\label{t:mglifetimes2}
\end{table*}

\begin{table*}
\caption{Oscillator strengths for Mg I lines}
\begin{tabular}{lcccc}
\hline  
Wavelength &  transition   & J'-J" & log $gf$  & log $gf$  \\
(\AA)      &               &       & adopted &   FF    \\
\hline  
 4057.51 & $3p^1P^0-8d^1D$ & 1-2 & $-$0.901 & $-$0.89 \\
 4167.28 & $3p^1P^0-7d^1D$ & 1-2 & $-$0.752 & $-$0.71 \\
 4703.00 & $3p^1P^0-5d^1D$ & 1-2 & $-$0.471 & $-$0.38 \\
 4730.04 & $3p^1P^0-6s^1S$ & 1-0 & $-$2.389 & $-$2.39 \\
 5167.33 & $3p^3P^0-4s^3S$ & 0-1 & $-$0.952 &         \\
 5172.70 &                 & 1-1 & $-$0.324 &         \\
 5183.62 &                 & 2-1 & $-$0.102 &         \\
 5528.42 & $3p^1P^0-4d^1D$ & 1-2 & $-$0.522 & $-$0.35 \\
 5711.10 & $3p^1P^0-5s^1S$ & 1-0 & $-$1.729 & $-$1.54 \\
 6318.71 & $4s^3S-6p^3P^0$ & 1-2 & $-$1.945 & $-$1.97 \\
 6319.24 &                 & 1-1 & $-$2.165 & $-$2.20 \\
 7657.61 & $4s^3S-5p^3P^0$ & 1-2 & $-$1.243 & $-$1.28 \\
 8209.85 & $3p^1P^0-8d^1D$ & 2-1 & $-$2.107 & $-$2.07 \\
 8712.7  & $4p^3P^0-7d^3D$ & 2-~ & $-$1.002 & $-$1.09 \\
 8717.8  &                 & 1-~ & $-$0.772 & $-$0.86 \\
 8923.6  & $4s^1S-4p^1P^0$ & 0-1 & $-$1.659 & $-$1.65 \\
 9432.7  & $4p^3P^0-6d^3D$ & 1-~ & $-$0.702 & $-$0.79 \\
10312.5  & $4p^1P^0-7d^1D$ & 1-2 & $-$1.558 & $-$1.52 \\
10953.3  & $4p^3P^0-5d^3D$ & 0-~ & $-$0.855 & $-$0.86 \\
11033.6  & $3d^3D-6p^3P^0$ & ~-1 & $-$2.059 & $-$2.10 \\
11522.3  & $4p^1P^0-6d^1D$ & 1-2 & $-$1.629 & $-$1.61 \\
12417.9  & $4p^3P^0-6s^3S$ & 0-1 & $-$1.587 & $-$1.63 \\
12423.0  &                 & 1-1 & $-$1.117 & $-$1.16 \\
12433.4  &                 & 2-1 & $-$0.897 & $-$0.94 \\
15879.5  & $3d^3D-5p^3P^0$ & ~-2 & $-$1.134 & $-$1.17 \\
15886.3  &                 & ~-1 & $-$1.354 & $-$1.39 \\
21458.9  & $5s^1S-6p^1P^0$ & 0-1 & $-$1.318 & $-$1.30 \\ 
\hline  
\end{tabular}
\label{t:mggf}
\end{table*}

\appendix{Appendix. Oscillator strengths for Mg I lines}

We found that the current status of oscillator strengths for lines of Mg~I is
not fully satisfactory. The values most used in the astronomical literature are
results of theoretical computations by Froese-Fischer (1975). However, there are
various more recent experimental evaluations of lifetimes of the levels relevant
for several optical transitions that can be used to improve these theoretical
estimates (Chantepie et al., 1989; J\"onsson et al. 1984; Kwiatkowski et al.
1980; Andersen et al. 1972; Schaefer 1971). Moreover, sophisticated theoretical
computations have been presented by Chang et al. for both lifetimes (Chang 1990a)
and oscillator strengths (Chang 1990b), for both singlet and triplet levels and
transitions. We used these various estimates to produce new values for the $gf$'s
of some Mg I lines. To this purpose, we coupled lifetimes (see
Table~\ref{t:mglifetimes1} and \ref{t:mglifetimes2}), given by a weighted
average of the mentioned experimental values, with branching ratios obtained
from theoretical oscillator strengths (see Table~\ref{t:mggf}).

While for most lines these oscillator strengths are very close to those of
Froese-Fisher (the mean difference from 24 lines is $0.00\pm 0.01$~dex, with an
r.m.s. of 0.07 dex for individual lines), for a few lines often used in the
analysis of metal-poor stars (like those at 4703, 5528 and 5711~\AA) the
oscillator strengths of Table~\ref{t:mggf} are smaller by as much as 0.19 dex.
This may cause both larger scatter and significant systematic errors in the
average Mg abundances in these stars.

The solar Mg abundance obtained using these $gf$'s, the appropriate
collisional damping parameters, and equivalent widths from Lambert \& Luck
(1978) is $\log n(Mg)=7.52\pm 0.05$\ when using the Holweger \& M\"uller
(1974) solar model atmosphere, and $\log n(Mg)=7.43\pm 0.06$\ when using the
Kurucz (1994) model atmosphere with the overshooting option switched off. Here
the error bars are the standard deviations of abundances from individual lines
about the mean value. These abundances are to be compared with the meteoritic
value of $\log n(Mg)=7.58\pm 0.02$\ (Anders \& Grevesse 1989). Note that
Asplund (2000) suggested that all meteoritic abundances may be overestimated
by 0.04 dex, due to a new Silicon abundance, lower than the value used by
Anders \& Grevesse. In view of these uncertaintes, the agreement between the
photospheric and meteoritic abundances is quite good when using the Holweger
\& M\"uller model atmosphere.

\end{document}